\documentclass[utf8]{FrontiersinHarvardPDF-NoLogo} 

\usepackage[onehalfspacing]{setspace}
\usepackage{lineno,microtype,subcaption,url,hyperref}

\def\keyFont{\fontsize{8}{11}\helveticabold }
\def\firstAuthorLast{Loredo {et~al.}} 
\def\Authors{Thomas J. Loredo\,$^{1,*}$ and Robert L. Wolpert\,$^{2}$}
\def\Address{$^{1}$Cornell Center for Astrophysics and Planetary Science (CCAPS) and Department of Statistics and Data Science, Cornell University, Ithaca NY, USA\\
$^{2}$Department of Statistical Science, Duke University, Durham NC, USA  }
\def\corrAuthor{Thomas Loredo}

\def\corrEmail{loredo@astro.cornell.edu}

\newcommand{\bi}{\begin{itemize}}
\newcommand{\ei}{\end{itemize}}
\newcommand{\bti}{\begin{titemize}}
\newcommand{\eti}{\end{titemize}}
\newcommand{\bqs}{\begin{quote}\small}
\newcommand{\eqs}{\end{quote}}
\newcommand{\bqt}{\begin{quote}\tiny}
\newcommand{\eqt}{\end{quote}}

\newcommand{\bc}{\begin{center}}
\newcommand{\ec}{\end{center}}

\newcommand{\be}{\begin{equation}}
\newcommand{\ee}{\end{equation}}
\newcommand{\bae}{\begin{align}}
\newcommand{\bea}{\begin{eqnarray}}
\newcommand{\eea}{\end{eqnarray}}
\def\bal#1\eal{\begin{align}#1\end{align}}

\newcommand{\bsplit}{\begin{split}}
\newcommand{\esplit}{\end{split}}
\def\bes#1#2\ees{\begin{equation}\label{#1}\begin{split}#2\end{split}\end{equation}}

\newcommand{\ceqn}[1]{equation~\eqref{#1}}

\newcommand{\cfig}[1]{Fig.~\ref{#1}}

\newcommand{\dif}{\mathop{}\!\mathrm{d}}

\newcommand{\ctxt}{\mathcal{C}}
\newcommand{\scond}[1]{\qquad||\;{#1}}  

\newcommand{\pmf}{{PMF}}
\newcommand{\pdf}{{PDF}}

\newcommand{\prob}{P}

\newcommand{\Dobs}{D_{\hbox{\tiny obs}}}

\newcommand{\like}{\mathcal{L}}

\newcommand{\eps}{\epsilon}

\newcommand{\normal}{\mathop{\mathrm{N}}}

\begin{document}
\onecolumn
\firstpage{1}

\title[Bayesian inference]{Bayesian inference: More than Bayes's theorem} 

\author[\firstAuthorLast ]{\Authors} 
\address{} 
\correspondance{} 

\extraAuth{}

\maketitle

\begin{abstract}


Bayesian inference gets its name from \emph{Bayes's theorem}, expressing posterior probabilities for hypotheses about a data generating process as the (normalized) product of prior probabilities and a likelihood function.
But Bayesian inference uses all of probability theory, not just Bayes's theorem.
Many hypotheses of scientific interest are \emph{composite hypotheses}, with the strength of evidence for the hypothesis dependent on knowledge about auxiliary factors, such as the values of nuisance parameters (e.g., uncertain background rates or calibration factors).
Many important capabilities of Bayesian methods arise from use of the law of total probability, which instructs analysts to compute probabilities for composite hypotheses by \emph{marginalization} over auxiliary factors.
This tutorial targets relative newcomers to Bayesian inference, aiming to complement tutorials that focus on Bayes's theorem and how priors modulate likelihoods.
The emphasis here is on marginalization over parameter spaces---both how it is the foundation for important capabilities, and how it may motivate caution when parameter spaces are large.
Topics covered include the difference between likelihood and probability, understanding the impact of priors beyond merely shifting the maximum likelihood estimate, and the role of marginalization in accounting for uncertainty in nuisance parameters, systematic error, and model misspecification. 

\tiny
 \keyFont{ \section{Keywords:} Astrostatistics, Bayesian methods, Poisson distribution, nuisanace parameters, systematic error, likelihood, marginalization} 
\end{abstract}

\section{Introduction}

The Bayesian approach to statistical inference and other data analysis tasks gets its name from \emph{Bayes's theorem} (BT).
BT specifies that a posterior probability for a hypothesis concerning a data generating process may be computed by multiplying a prior probability and a likelihood function (and normalizing):
\be
\text{posterior} \propto \text{prior} \times \text{likelihood}.
\label{BT-text}
\ee
Tutorials on Bayesian methods often focus on how BT adjusts the likelihood function to account for base rates of hypotheses about members of a population,  using simple examples like binary classification (e.g., disease/no-disease; guilty/not-guilty; star/galaxy) based on binary diagnostic data (e.g., a positive or negative test or evidentiary result). 
At the \emph{iid2022: Statistical Methods for Event Data} meeting that is the topic of this special issue of \emph{Frontiers}, a charming and insightful tutorial along these lines was presented by Torsten En{\ss}lin; see his ``Bayes Basics'' presentation at the \href{https://sites.google.com/uah.edu/iid2022/}{meeting's website}.


In this tutorial contribution to the issue we address readers familiar with (and perhaps already using) Bayesian methods, to make the case that a focus on Bayes's theorem risks overlooking a more fundamental and crucial aspect of Bayesian inference.
We consider a key distinguishing feature of Bayesian inference to be is its use of the \emph{law of total probability} (LTP), directing one to \emph{compute probabilities for composite hypotheses by marginalizing over the hypothesis or parameter space}.

We do not consider this a new or controversial insight.
The first author came to appreciate it as a student of the work of \citet{J61-TOP}, \citet{J03-PTLOS}, \citet{Z71-BayesEcon}, and other mid-20th-century pioneers of modern Bayesian inference, and through the work of the second author and Jim Berger (\citet{BW88-LikePrinc}, especially \S\S~3.5 and 5.3). 
We have highlighted this in the past (e.g., in \citet{L92-Promise,L99-BayesComp,L13-BackwardLook}, \citet{BLW99-IntegLike}, and in lectures at the Penn State \href{https://sites.psu.edu/astrostatistics/past-programs/}{\emph{Summer School in Statistics for Astronomers}}).
But we feel this aspect of Bayesian analysis is underappreciated, especially by newcomers and non-experts.
With the growing importance of high-dimensional models in statistics and machine learning (ML) there is new motivation for highlighting it.
In particular, ML relies heavily on \emph{optimization over large parameter spaces}, where, from a Bayesian perspective, marginalization may instead be the right operation.
Recent research demonstrates that the performance of some ML models can be significantly enhanced by replacing optimization with marginalization (even approximately), particularly in settings were good uncertainty quantification is important.
Andrew Wilson's Bayesian machine learning group has done particularly notable work in this direction \citep{W20-BayesDL,WI20-BayesDL}.
It seems to us the crucial importance of the LTP and marginalization deserves amplification, particular for newcomers to Bayesian methods.

In this tutorial, we start by establishing notation and informally reviewing BT, the LTP, and the Bayesian interpretation of probability.
Then we explore the role of marginalization in Bayesian inference in the context of the following topics:
\bi
\item \emph{Likelihood vs. probability:}
We review Fisher's introduction of likelihood as a complement to probability, and the use of BT to ``flip the conditional'' and create a posterior distribution from a likelihood function.
Newcomers to Bayesian inference who use the term ``likelihood of the data'' will find a corrective here.




\item \emph{Priors are not (merely) penalties:}
A prior distribution does more than just shift the peak of the likelihood function; it converts the likelihood to a quantity that can be meaningfully integrated.
In even modest-dimension spaces, so-called ``curses of dimensionality'' and related ideas imply that integrated probability can pile up in unexpected ways,  away from the peak of the likelihood function.

\item \emph{Marginalization vs.\ optimization over nuisance parameters:}
Most data analysis problems in astronomy rely on models that include both parameters of direct scientific interest, and \emph{nuisance parameters}---parameters (e.g., describing backgrounds) that are necessary for linking the interesting parameters to the data.
Bayesian methods marginalize (average) over nuisance parameters to account for their uncertainty.
We contrast marginalization with a popular alternative approach relying on optimization instead.
We also discuss so-called \emph{measurement error problems} (a statistics term of art), where the differences between marginalization and optimization can be amplified.

\item \emph{Marginalization and systematic error:}
We briefly describe recent and ongoing work using marginalization to describe and propagate systematic error in settings where standard ``propagation of error'' (the statisticians' ``delta method'') fails or is inapplicable.
An appendix provides more details on some of this work.
\ei

\section{Notation and basic concepts}

We adopt notation similar to that of Jeffreys, Cox, and Jaynes, who view Bayesian inference as a generalized logic \citep{J61-TOP,C46-ProbAxioms,J03-PTLOS}.
Whereas deductive logic provides a calculus for truth and falsity in settings where we can reason with certainty (e.g., with truth and falsity represented by 1 and 0 in Boolean algebra), probability theory provides a calculus for degrees of entailment, or argument strength, in settings where we must quantify uncertainty (with probabilities taking values over the real interval, $[0,1]$).
We use $\prob(A|B)$ to denote the probability that the truth of statement $A$ follows from taking statement $B$ to be true (whether $B$ is known to be true or just assumed to be true).
We call this the conditional probability for $A$ given $B$.
Interpreting probability as a measure of entailment or argument strength means that all probabilities are necessarily conditional.
The argument of a probability symbol is the whole expression inside the parentheses---$A|B$ here---understood as the statement that the truth of $A$ follows from the truth $B$.
In logic, such a statement is called an \emph{argument}, so we may say that the argument of a (Bayesian) probability symbol is, well, an argument.
The conditioning statements comprise the premises for the argument (the ``givens'').

We will often be computing a collection of probabilities that share common conditions that will not be questioned (i.e., they never appear to the left of the conditioning bar), at least for the duration of a specific calculation.
We often denote such conditioning information by $\ctxt$, the \emph{context} for the calculation.
Some contextual information may represent secure knowledge (e.g., basic physical principles that firmly justify some probability distributions appearing in an analysis).
Other contextual information may represent less secure, provisional assumptions, adopted ``for the sake of the argument,'' to enable us to compute probabilities required in the course of an analysis; such assumptions ideally should be reassessed in a later stage of analysis.%
\footnote{
John Tukey---best known for his work on the fast Fourier transform and exploratory data analysis---observed: ``No body of data tells us all we need to know about its own analysis'' \citep{T77-EDA}.
Data do not ``speak for themselves'' in scientific arguments; data analysis considers the implications of data within some context.
In this sense, data analysis (Bayesian or frequentist) is necessarily subjective and provisional, and a virtue of the Jeffreys/Cox/Jaynes notation is its recognition that all probabilities are conditional, at least depending on the context, $\ctxt$ (whose components should be explicitly identified).
For further discussion of subjectivity in statistical data analysis we recommend perspective papers by \cite{BB88-StatAnIllusObjectivity}, \cite{L00-PhilStat}, and \cite{GH17-BeyondSubjObjStat} (the latter including wide-ranging remarks by many invited discussants).}


Complex statements may be built by combining simpler statements.
For example, we use $(A,B)$ to denote ``$A$ and $B$ are both true,'' and $(A\lor B)$ to denote ``$A$ is true, or $B$ is true, or both.''
Two important basic rules of probability are the conjunction rule (``and'' or ``product'' rule):
\be
\begin{split}
\prob(A, B|\ctxt) 
 &= \prob(A|\ctxt) \, \prob(B|A, \ctxt)\\
 &= \prob(B|\ctxt) \, \prob(A|B, \ctxt),
\end{split}
\label{and-rule}
\ee
and the disjunction rule (``or'' or ``sum'' rule):
\be
\prob(A\lor B|\ctxt) = \prob(A|\ctxt) + \prob(B|\ctxt)
 - \prob(A, B|\ctxt).
\label{or-rule}
\ee
We presume the reader is familiar with these basics; they are here to establish notation and terminology.

We often compute probabilities for collections of statements labeled by an integer or a real-valued parameter (or vector of parameters).
In such cases we use a lower-case $p$ (or another letter) to denote the appropriately labeled probabilities.
For discrete cases, the probabilities comprise a \emph{probability mass function} (\pmf), such as $p_i = \prob(A_i|\ctxt)$ (for $i$ an integer over some specified range).
For continuous cases, we instead use a \emph{probability density function} (\pdf), defined so that for a parameter $\theta$ the probability assigned to an interval of size $d\theta$ is $p(\theta)\,d\theta = \prob(\theta_* \in [\theta, \theta + d\theta]|\ctxt)$, where $\theta_*$ is the (uncertain) true value of the parameter.
So the argument of a probability symbol is an \emph{argument} (comprised of statements comprising premises and a conclusion), and the argument or index of a \pmf\ or \pdf\ symbol is a \emph{number}.


When conditioning information is common to all probability symbols in a formula it can be distracting to display it explicitly.
We sometimes suppress symbols for common conditions for clarity, when the conditions are clear from the context.
When we do want to display a common dependence, we adopt clever notation introduced by John Skilling (the ``Skilling conditional''), showing the shared information with a double bar beside the equations, as here:
\be
\prob(A, B) 
  = \prob(A) \, \prob(B|A) \scond{\ctxt}.
\label{SCond}
\ee

In what follows we will focus on Bayesian \emph{inference}---computing probabilities for hypotheses of interest given data and other information, including modeling assumptions.
Bayesian inference is the core of \emph{Bayesian data analysis} (BDA), comprising the use of Bayesian probability for a broader variety of tasks, including making decisions (Bayesian decision theory), designing experiments (Bayesian experimental design), and addressing tasks where there is not enough structure for formal inference (e.g., Bayesian model checking and Bayesian exploratory data analysis).
Marginalization plays a key role in all of these data analytical activities.
The book \emph{Bayesian Data Analysis} \citep{G+14-BDA} provides comprehensive coverage of many topics comprising BDA.
An emerging term of art, \emph{Bayesian workflow}, refers to best practices for integrating diverse Bayesian techniques (including methods fusing Bayesian and frequentist considerations) into robust data analysis pipelines that include model checking and refinement steps.
\cite{G+20-BayesWorkflow} provide a broad discussion of Bayesian workflow; \cite{TGE18-ProperBayesAstro} and \cite{E+23-BayesGuidance} discuss key components of sound Bayesian workflow in astrostatistics.


\section{Bayes's theorem and interpreting probability}

If we equate the two factorizations on the right hand side of \ceqn{and-rule} and solve for $\prob(A|B,\ctxt)$, we get Bayes's theorem,
\be
\prob(A|B) = \frac{\prob(A)\,\prob(B|A)}{\prob(B)} \scond{\ctxt},
\ee
provided that $\prob(B)\ne 0$.%
\footnote{In settings where the alternatives are labeled by a continuous parameter, $\theta$, $B$ would represent a statement about the value of $\theta$, i.e., that it is positive, or lies in a specified interval.
When the probability $\prob(B)$ can be computed using a continuous PDF, $p(\theta)$, the probability that $\theta$ takes a specified value $\theta_*$, $P(B: \theta = \theta_*)$, formally vanishes (it is equal to $p(\theta)\dif\theta$ in the limit as $\dif\theta\rightarrow 0$).
In such settings, it can be possible to condition on $\theta = \theta_*$ even though $P(B) = 0$, by considering a limit, or more generally, via measure-theoretic arguments.
See \cite{MR10-SDDR} for discussion of this and its potential relevance for Bayesian model comparison.}
We can view BT as showing how adding statements to the premises in an argument should change the degree of entailment for the conclusion.
Here the initial argument is that $A$ follows from $\ctxt$ (the argument of the first factor in the numerator), and the argument of the probability on the left hand side is asserts that $A$ follows from the combined statement $(B,\ctxt)$, i.e., $B$ has been added to the premises.

To get BT in a form useful for data analysis, let $H_i$ denote statements asserting rival hypotheses specifying a data generating process (DGP), indexed by $i$; let $\Dobs$ denote a statement asserting the values of observed data; and let $\ctxt$ denote all other information at hand, including a description of how the hypotheses and data are connected.
Rewrite BT, taking $A=H_i$ and $B=\Dobs$, giving
\be
\prob(H_i|\Dobs) = \frac{\prob(H_i)\,\prob(\Dobs|H_i)}{\prob(\Dobs)} \scond{\ctxt}.
\label{BT-PMF}
\ee
With these choices for the statements, the factors in this equation have names:
\bi
\item $\prob(H_i|\Dobs)$ is the \emph{posterior probability} for hypothesis $H_i$ (given $\Dobs$).
Considered as a function of $i$ it is a posterior \pmf.
``Posterior'' here refers to ``\emph{after} taking into account the observed data.''

\item $\prob(H_i)$ is the \emph{prior probability} for $H_i$, also a \pmf\ when considered as a function of $i$.
``Prior'' here refers to ``\emph{prior} to accounting for the observed data.''

\item $\prob(\Dobs|H_i)$ is the \emph{likelihood} for hypothesis $H_i$, or, considered as a function of $i$, the \emph{likelihood function}.
The term ``likelihood'' is meant to indicate it is not a \emph{probability} for $H_i$, since $H_i$ is on the wrong side of the bar (in particular,  $\sum_i \prob(\Dobs|H_i)$ need not equal unity).
We discuss the relationship between probability and likelihood further in \S~\ref{sec:LP}.

\item $\prob(\Dobs)$ is the \emph{prior predictive probability}, the probability with which one would predict the observed data without specifying which of the $H_i$ is true.
\ei
If we are considering a continuum of hypotheses labeled by a continuous parameter $\theta$, then one may derive an analogous result involving \pdf{s}:
\be
p(\theta|\Dobs) = \frac{p(\theta)\,p(\Dobs|\theta)}{p(\Dobs)} \scond{\ctxt}
\label{BT-PDF}
\ee
(where we have mildly overloaded our notation, using $\theta$ and $\Dobs$ to 
denote \emph{variables} for the scalar or vector parameter and data quantities 
when to the left of the bar, and \emph{statements} asserting values for those quantities when to the right).
The factors have analogous names, e.g., $p(\theta|\Dobs)$ is the posterior \pdf\ for $\theta$.

There is nothing controversial in BT as an uninterpreted mathematical result.
But a problem arises in its use for data analysis if one insists on adopting a frequentist interpretation of probability.
Such interpretations assert that probabilities can be meaningfully assigned and manipulated only for statements about quantities that take on diverse values in some kind of \emph{replication} setting; such statements are called \emph{events}.%
\footnote{A statement $D_{\rm\tiny obs}$ asserting values for observed data---e.g., ``The number of heads in 10 flips was 7'', or ``The photon count in bin 1 is 11, the photon count in bin 2 is 15\dots''---can typically be considered an event in this sense, interpreting the sampling distribution as a distribution for repeated sampling.}
In such settings, the probability for an event is the fraction of the time the event happens (the statement is true) among replications, as the number of replications tends to infinity.%
\footnote{This idea is notoriously difficult to define rigorously; arguably there is not yet a sound definition.
For an exceptionally accessible overview of the difficulties see \citet{DS18-TenIdeas}.}
The frequentist view is significantly restrictive.
It allows us to say, ``the probability for heads is $1/2$,'' interpreted as a property of an \emph{ensemble} of flips of a coin.
But it does not allow us to say, ``the probability for heads \emph{on the next flip} is $1/2$,'' if we are about to flip a coin just once.

More importantly, the frequentist interpretation does not permit assigning or computing probabilities for hypotheses about fixed but uncertain properties of a physical system.
For example, in the 1800s Laplace famously used Bayesian probability (the historically original notion of probability) to analyze detailed but noisy observations of Saturn's orbit to estimate the ratio of Saturn's mass to the Sun's mass.
He estimated the ratio to be within $\pm 1\%$ of $2.847\times 10^{-4}$, with a probability of 0.99991.
(The modern convention would fix the probability at some convenient target like 0.95 and find the associated range; Laplace did the reverse.
The current estimate is $2.858\times 10^{-4}$, comfortably inside Laplace's range.)
Such a probability makes no sense if one adopts a frequentist interpretation.
The true mass of Saturn either is or is not in that interval, and it would be either in it or outside of it for every replication of the observations.
The frequentist probability for Saturn's mass being in that interval is either 0 or 1---but we have no way of knowing which it is.

\begin{figure}[t]
\begin{center}
\includegraphics[width=0.45\textwidth]{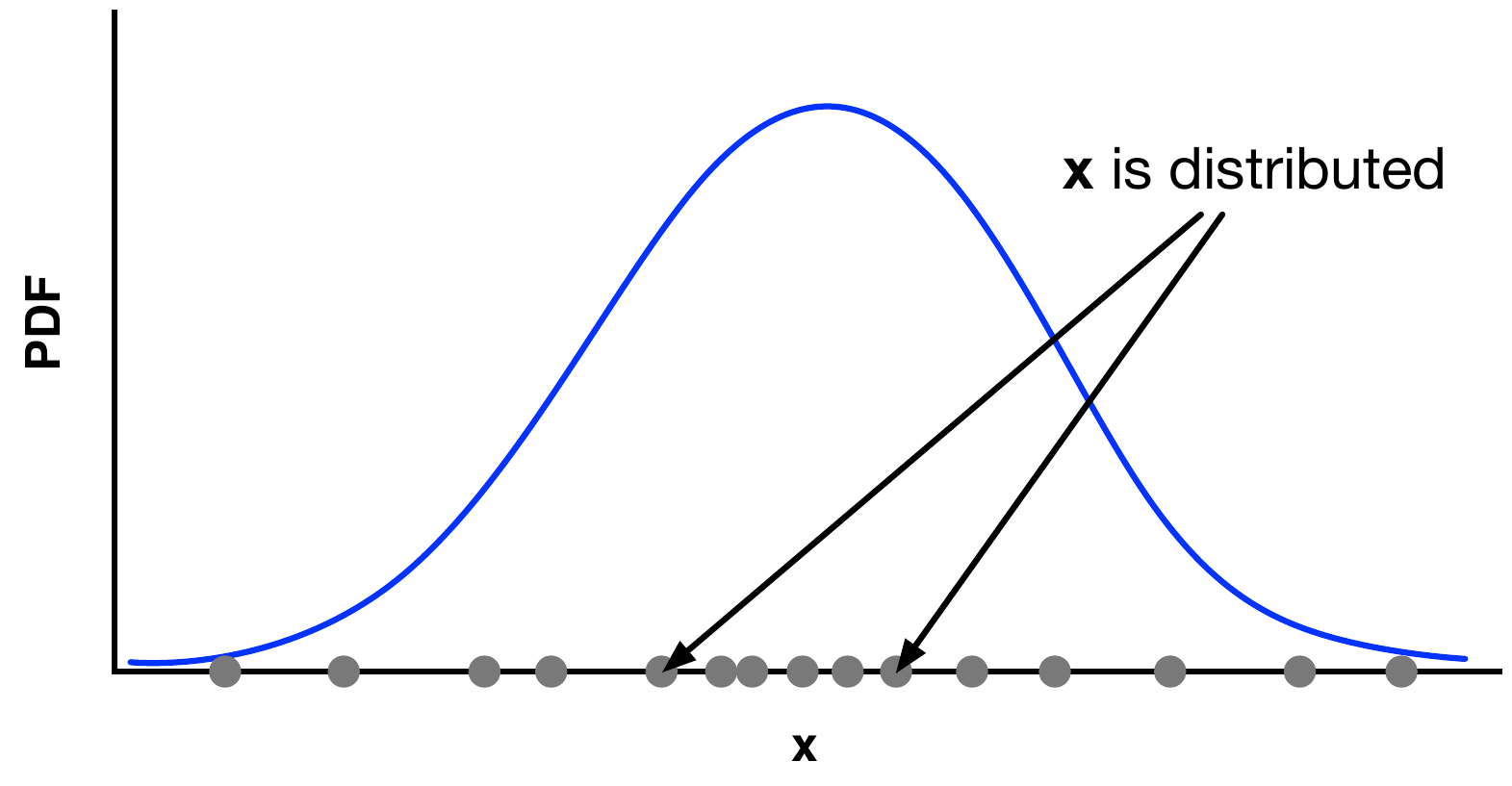}\qquad
\includegraphics[width=0.45\textwidth]{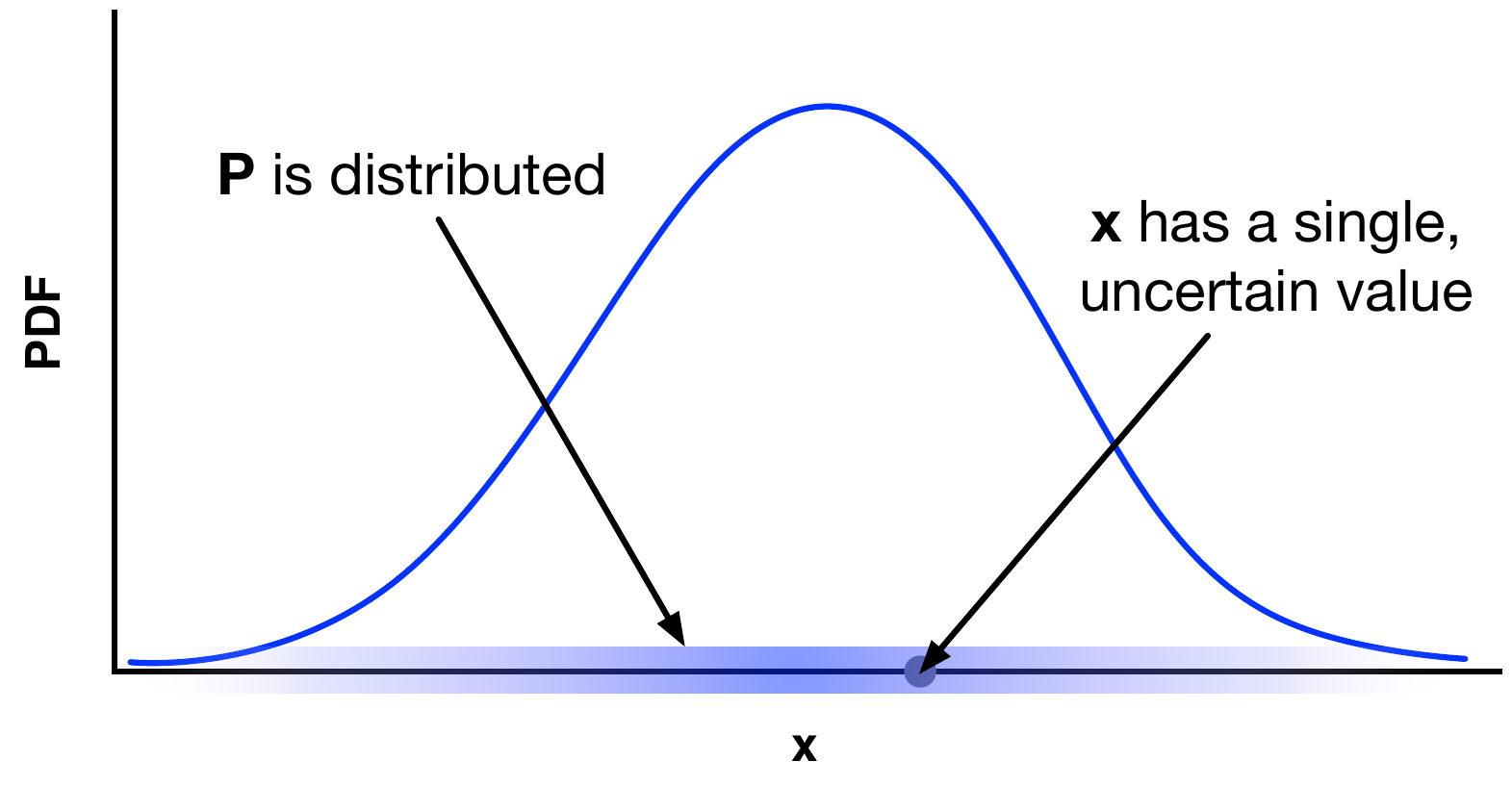}
\end{center}
\caption{Frequentist and Bayesian interpretations of a \pdf\ (blue curve) for a real-valued quantity $x$.
\emph{Left:} Frequentist interpretation as a (limiting) histogram of values $x$ takes across replications; the values of $x$ are distributed according to the \pdf.
\emph{Right:} Bayesian interpretation as $x$ taking a single fixed (but uncertain) value in the case at hand (dot on the $x$ axis), with probability distributed over the possible values (depicted via the shading along the $x$ axis).}
\label{fig:PDFs}
\end{figure}

One sometimes hears a complaint from a scientist only familiar with the frequentist interpretation that they can follow the Bayesian math, but stumble at the idea that a parameter value should be considered to be ``random,'' a random variable taking on a distribution of values.
This reflects an error in interpretation.
\cfig{fig:PDFs} is our attempt at depicting the different interpretations for a \pdf\ for some real-valued quantity $x$ (the interpretations differ regardless of whether $x$ refers the value of a parameter or a datum).
The left panel depicts the frequentist interpretation of $p(x)$; the \pdf\ describes an infinite collection of replications, across which $x$ takes on many values, whose limiting histogram is $p(x)$.
The right panel depicts the Bayesian interpretation; the \pdf\ describes uncertainty about the value of $x$ in a single case-at-hand by distributing probability over the values $x$ may take (depicted by the shading along the $x$ axis).
In the frequentist interpretation, it's the $x$ in $p(x)$ that is distributed (across replications).
In the Bayesian interpretation, it's the $p$ in $p(x)$ that is distributed (across possible values $x$ may take in the case at hand).
A Bayesian \pdf\ is analogous to a matter density, $\rho(x)$, in classical mechanics; it is the matter that is distributed, not values of $x$.

\section{The law of total probability and marginalization}
\label{sec:LTP}

BT followed from the conjunction (``and'') rule.
The LTP follows from application of both the conjunction and disjunction (``or'') rules to problems where a statement of interest may be true or false depending on what is true about auxiliary details in the problem.
As simple example, consider rolling a die, and a hypothesis about the outcome of a single roll, $H =$ ``The number of dots on the top face is a prime.''
Let the auxiliary detail be the actual number that comes up, with $B_i =$ ``The side with $i$ dots is up.''
Here $H$ is true when any of $B_2$, $B_3$, or $B_5$ is true, and false otherwise.
Note also that only one of the $B_i$ statements may be true.
We say that $H$ is a \emph{composite hypothesis}: there are multiple ways it may be true or false, given the nature of the problem (the context).
Here the $B_i$ comprise an \emph{exclusive, exhaustive set of statements}---a collection of statements such that our contextual information asserts one of them must be true, but only one. 
\citet{D21-ThinkBayes} dubs such a set a \emph{suite}, terminology we like that is not yet widely used.

In such settings, the joint probability (``and'') for $H$ and one of the $B_i$ may be factored as $P(H,B_i) = P(H)\, P(B_i|H)$.
Now sum this over $i$, noting that $P(H)$ does not depend on $i$ and so may be taken out of the sum:
\be
\sum_i P(H,B_i) = P(H) \sum_i P(B_i|H) = P(H)  \scond{\ctxt},
\ee
where for the final equation we used repeated application of the ``or'' rule and the suite property to show that $\sum_i P(B_i|H,\ctxt)$ is unity.
We thus have the LTP:
\bea
\begin{split}
P(H)
  &= \sum_i P(H,B_i) \\
  &= \sum_i P(B_i)\, P(H|B_i)  \scond{\ctxt},
\end{split}
\label{LTP}
\eea
where the last line uses the alternative factorization of $P(H,B_i)$.
The summing over auxiliary details is called \emph{marginalization}, after the practice of collecting values of a joint probability in a table and listing sums across rows and columns in the margins of the table. 
The LTP is a kind of ``decomposition/recomposition'' result that is used in two ways.

First, if we have a problem where both $H$ (or a set of alternatives, $H_j$) and $B_i$ are present from the start, so $P(H,B_i)$ is available, the top line of \ceqn{LTP} tells us how to compute the probability for $H$ if it alone is of interest: sum the joint over the ways $H$ may be true.

Second, if we have a problem where initially only $H$ is present but we do not immediately see how to compute or assign $P(H)$, the second line of \ceqn{LTP} suggests doing something reminiscent of a basis expansion:
identify some additional detail that, if known, would let you compute the probability for $H$; that is, find a suite, $B_i$, so that $P(H|B_i)$ is known.
Then sum over the possible choices of auxiliary details, weighting the terms by the probabilities for each choice.
This has been called ``extending the conversation'' \citep{L78-BayesApproach}.
Harvard statistician Joseph Blitzstein offers a different term in one of his YouTube lectures (see ``\href{https://youtu.be/fDcjhAKuhqQ?t=958}{Statistics 110, Lecture 6: Monty Hall, Simpson's Paradox},'' slightly paraphrased here):
\begin{quote}
In most mathematical subjects, if you have a problem and you're stuck,
saying ``I wish I knew this or that'' doesn't help you.
In probability theory, thinking ``I wish I knew this'' gives you a hint at what you should condition on.  Then you condition on it, act \emph{as if} you did know it, and then average over those possibilities.

\noindent
I didn't name the law of total probability, but if I had,
I would have just called it \emph{wishful thinking}: what do I wish I knew?
\end{quote}

In continuous-parameter settings where uncertainty may be quantified by PDFs, the LTP instructs us to \emph{integrate} over those parameters specifying choices of auxiliary details.

In a typical Bayesian data analysis application, one will begin by using BT to define a posterior distribution, but then use that distribution by applying the LTP to address diverse tasks.
Here are several such tasks:

\begin{itemize}
\item \emph{Normalizing posterior \pdf{s}:}
The posterior predictive probability in the denominator of BT in \ceqn{BT-PDF} appears to play the role of a normalization constant, in that it does not depend on $\theta$.
We can show this explicitly using the ``wishful thinking'' version of the LTP:
\be
p(\Dobs) = \int\dif\theta\, p(\theta)\,p(\Dobs|\theta).
\label{PriorPred}
\ee
Since the integrand is just the numerator of \ceqn{BT-PDF}, the LTP has both verified that the prior predictive probability is a normalization constant, and shown us how to compute it.
It is also the source of the more common name for this quantity, the \emph{marginal likelihood} (for the model as a whole), i.e., the weighted integral of the likelihood function.

\item \emph{Credible regions:} Calculating the probability in a {\em credible region}, $R$, for $\theta$ may be done similarly:
\begin{equation}
p(\theta \in R|\Dobs)
  = \int_R d\theta\, p(\theta \in R|\theta)\, p(\theta|\Dobs).
\label{HPDCR}
\end{equation}
There may be many regions that have a desired target amount of probability.
The smallest such region is typically unique, with density higher in $R$ than outside it, and it is called a \emph{highest posterior density} (HPD) credible region.

\item \emph{Marginalizing over nuisance parameters:}
Most real-life data analysis problems in astronomy have \emph{nuisance parameters}: parameters required for modeling the data, but not directly interesting.
Perhaps the most common such parameters are those describing backgrounds contaminating the measurement of a signal.
Let $\psi$ denote the interesting (signal) parameters, and $\eta$ the nuisance (background) parameters.
The uncertainty in $\psi$ (with the $\eta$ uncertainty fully propagated) is
quantified by the \emph{marginal posterior \pdf},
\begin{equation}
p(\psi |\Dobs) = \int d\eta\, p(\psi,\eta|\Dobs).
\label{marg-pdf}
\end{equation}
If the posterior is explored via posterior sampling (say, via MCMC), this integral may be approximated by simply making a histogram of the $\psi$ component of the posterior samples of $(\psi,\eta)$.
A popular way to handle nuisance parameters in astronomy and physics is via \emph{profile likelihood}, which optimizes rather than averages over the nuisance parameters.
We discuss the relationship between marginalization and optimization (profiling) at some length in \S~\ref{sec:marg}.

\item \emph{Propagation of uncertainty:}
Suppose we have computed a posterior \pdf\ for a model's parameters, $\theta$, but we are interested, not directly in $\theta$, but in a quantity $f = F(\theta)$ for known function $F(\theta)$.
This is the setting for standard ``propagation of errors'' techniques (known as the ``delta method'' in statistics), but these rely on approximations that presume the posterior is Gaussian.
We can propagate uncertainty more thoroughly and accurately using the LTP to compute a marginal posterior for $f$:
\be
\begin{split}
p(f|D,M) 
  &= \int d\theta\; p(f,\theta|\Dobs,M) 
      \;\,=\;\, \int \dif\theta\; p(\theta|\Dobs,M)\, p(f|\theta,M) \\
  &= \int d\theta\; p(\theta|\Dobs,M)\, \delta[f - F(\theta)].
\end{split}
\label{propgn}
\ee
If the posterior is explored via posterior sampling, this integral may be approximated by simply making a histogram of $F(\theta)$ evaluated over the posterior samples of $\theta$.

\item \emph{Prediction:}
In the context of a model with parameters $\theta$ for observed data $\Dobs$, where we want to make predictions about future data we may obtain (say, for experimental design purposes), we can use the LTP to compute a \emph{posterior predictive distribution}.
Denoting future data by $D'$, the posterior predictive PDF is
\begin{equation}
p(D'|\Dobs) = \int d\theta\, p(D'|\theta)\, p(\theta|\Dobs),
\label{post-predict}
\end{equation}
with the integration accounting for parameter uncertainty in the prediction, weighting possible choices of $\theta$ by the posterior \pdf.

\item \emph{Model comparison:}
To compare rival parametric models defined by contexts $M_i$ (each with parameters $\theta_i$), we compute posterior odds or Bayes factors.
These require computation of each model's {\em marginal likelihood},%
\footnote{Readers new to model comparison should note that in the astronomy literature and some machine learning literature it is popular to use the term \emph{evidence} for the marginal likelihood.
(\emph{Bayesian evidence} also appears, awkwardly implying there are Bayesian vs.\ non-Bayesian forms of evidence.)
We eschew this terminology.
The \emph{data} (and possibly additional contextual information) comprise the evidence, and in the Bayesian philosophy of science literature it has long been common to use $E$ and ``evidence'' where we have used $D_{\rm\tiny obs}$ and ``data.''
Unsurprisingly, ``evidence'' is used in the same way in the literature on statistics in jurisprudence.
Using ``evidence'' for marginal likelihood obscures that there are just two types of hypothesis-dependent quantities in Bayesian inference, the probability for a hypothesis (with the hypothesis to the left of the bar) and the likelihood for a hypothesis (with the hypothesis to the right of the bar).
The marginal likelihood works just like the more familiar parameter likelihood function; ``marginal'' merely specifies how it is computed.
Jack Good has noted that marginal likelihoods operate just like likelihood functions, calling them \emph{Bayesian likelihoods}.
He used \emph{weight of evidence} for the logarithm of the ratio of marginal likelihoods for two hypotheses (i.e., the logarithm of the \emph{Bayes factor}), providing an additive (rather than multiplicative) measure of how much the available evidence favores one hypothesis over the other; the terminology originates with Alan Turing \citep{G50-ProbWeighEv,G85-WtEvSurvey}.
}
\begin{equation}
p(\Dobs|M_i) = \int d\theta_i\, p(\theta_i|M_i)\, p(\Dobs|\theta_i,M_i)
  \scond{M_1\lor M_2\ldots}.
\label{marg-like}
\end{equation}
This is just the normalization constant for the posterior \pdf\ for a particular model's parameters.
This says that the likelihood for a model (as a whole, i.e., accounting for uncertainty in its parameters) is the average of
the likelihood function for that model's parameters.%
\footnote{The values of marginal likelihoods depend more sensitively on properties of prior distributions than do posterior PDFs for parameters; in particular, they are roughly inversely proportional to the prior ranges of parameters.
As a result, formal Bayesian model comparison and model averaging are best reserved for problems where the analyst can strongly motivate priors (especially prior ranges), or quantitatively study how results depend on properties of priors.
Good entry points to the literature on this (in both astronomy and statistics) include: \cite{V+99-BayesModelAvgTut,CG04-ModelUncert,C+07-BayesModelChoice,T12-CosmoBayesModelComp}.
For a Bayesian perspective on model selection relying on measures of anticipated out-of-sample predictive performance (vs.\ a prior prediction of the observed sample, as measured by marginal likelihood), see \cite{GHV14-InfoCriteriaBayes}.
}

\item \emph{Model averaging:}
Consider again a setting with rival parametric models with contexts $M_i$, but where all models share some interesting parameters, $\phi$, but supplement them with different sets of nuisance parameters, $\eta_i$.
An example from cosmology is estimating the curvature and size of the universe, accounting for uncertainty in the choice of cosmological models (e.g., whether dark energy is due to a cosmological constant or is instead evolving, thus possibly requiring additional parameters for the dark energy equation of state; see \citealt{VTS11-BayesModelAvgCosmo}).
In settings where no particular model is strongly preferred over its rivals, we would like to account for model uncertainty in estimation of $\phi$.
Let $\ctxt$ here denote an overall context, collecting all rival models.
Model averaging uses the LTP to propagate both model and nuisance parameter uncertainty as follows:
\be
\begin{split}
p(\phi|D,\ctxt)
  &= \sum_i p(M_i|D,\ctxt)\, p(\phi|D,M_i) \\
  &\propto \sum_i p(\Dobs|M_i) \int d\eta_i\, p(\phi,\eta_i|D,M_i),
\label{ModelAvg}
\end{split}
\ee
where for the last line we presume the models are considered equally probable a priori, so that the probability for each model is proportional to its marginal likelihood, $p(\Dobs|M_i)$.

\end{itemize}

This is not an exhaustive list, but surely is long enough to make the case for the powerful role the LTP plays in Bayesian inference.



\section{Likelihood vs.\ probability}
\label{sec:LP}

A \emph{parametric model} for data specifies how one may predict or simulate hypothetical data, $D$, as a function of a fixed-dimension parameter vector, $\theta$.
That is, a parametric model is a \emph{collection of sampling distributions}, a bivariate function of hypothetical values of the parameter, and hypothetical values of the data:%
\be
p(D|\theta) = f(D;\theta).
\label{SampDistns}
\ee
Each sampling distribution is a probability disribution \emph{for data}.

The likelihood function \emph{fixes} $D = \Dobs$, yielding a function only of $\theta$:
\be
\like(\theta)
  \equiv p(\Dobs|\theta)
  = f(\Dobs;\theta).
\label{like-def}
\ee
The likelihood function is \emph{not} a probability distribution for $\theta$ ($\theta$ is on the wrong side of the bar).
Strictly speaking, any positive multiple of $p(\Dobs|\theta)$ may serve as the likelihood function.
The multiplier may be constant or a function of $\Dobs$; such factors do not change the dependence of $\like(\theta)$ on $\theta$, which is what governs inference \citep{BW88-LikePrinc}.
Any such factor would also appear in the prior predictive probability, $p(\Dobs)$, and thus cancel in BT.

\begin{figure}[!t]
\begin{center}
\includegraphics[width=0.6\textwidth]{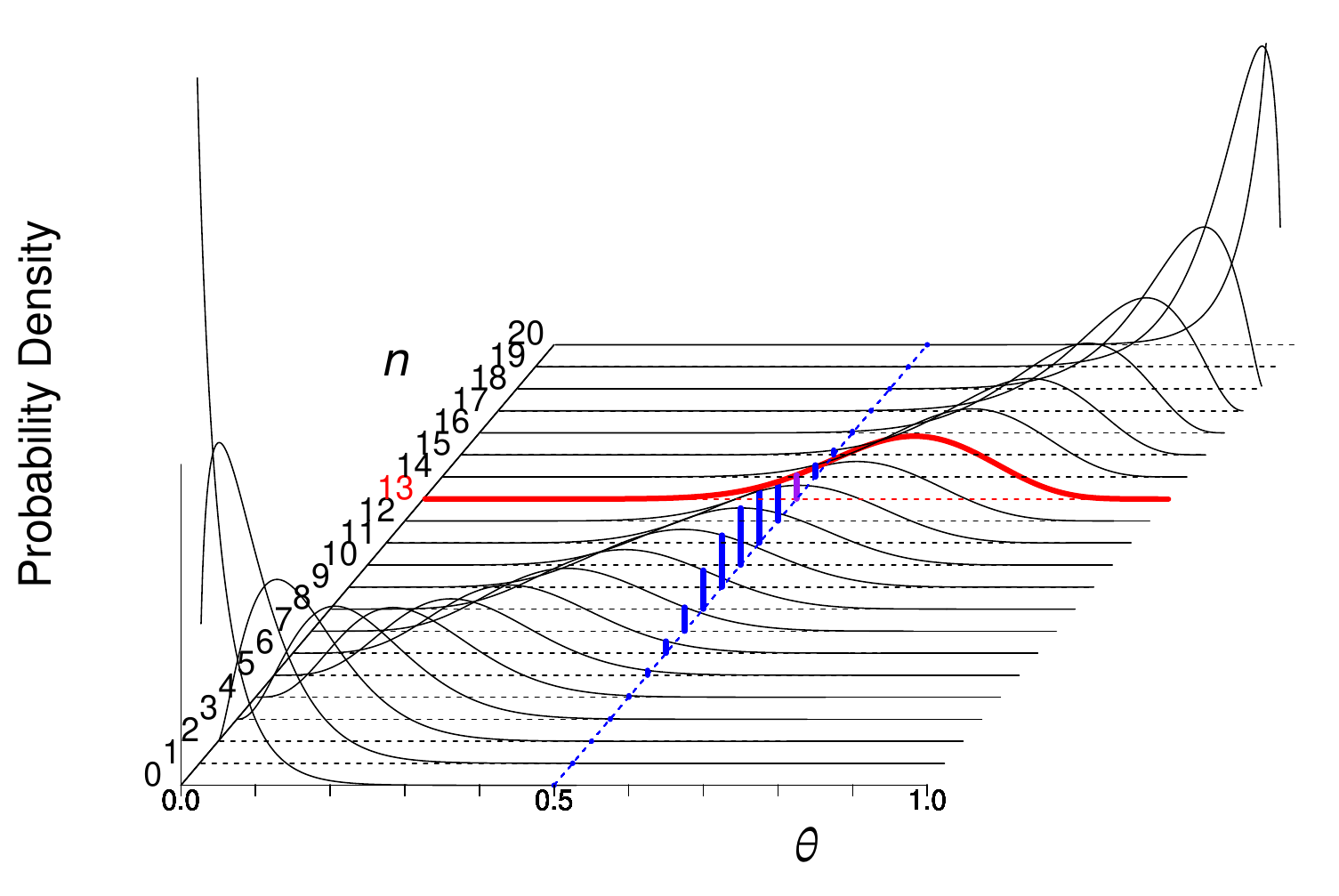}
\end{center}
\caption{The relationship between the sampling distribution and likelihood functions for estimation of the success probability parameter, $\alpha$, specifying a binomial distribution for the number of successes, $n$, expected in $N=20$ trials.
The blue histogram shows the binomial \pmf\ as a function of $n$ for one particular choice of $\theta$ (0.5).
The black curves show potential likelihood functions corresponding to different numbers of successes, $n$.
Once an observed value, $n=13$, is available, the actual likelihood is identified as the red curve.}
\label{fig:binomlike}
\end{figure}

In \cfig{fig:binomlike} we depict the relationship between the sampling distribution and the likelihood function for estimation of the probability of a binary outcome from data counting the number of successes in a sequence of trials or tests---think of counting the number of heads in flips of coins, or the number of stars in a sequence of star/galaxy separation tests in image data.
This is a case where the sample space is discrete (and so the sampling distribution is a \pmf), and the parameter space is continuous (so the likelihood function is continuous).
The blue histogram shows the binomial distribution giving the probability for seeing $n$ successes in $N$ total trials with a success probability specified by the real-valued parameter $\theta$,
\be
p(n|\theta,\ctxt) = \frac{N!}{n! (N-n)!} \theta^n (1-\theta)^{N-n},
\label{binom}
\ee
where the context includes specification of the total number of trials, $N$.
The black curves show potential likelihood functions corresponding to different numbers of successes, $n$.
Once $n=13$ is observed, the particular curve highlighted in red is identified as the likelihood function relevant for inference.

\begin{figure}[!t]
\begin{center}
\includegraphics[width=0.95\textwidth]{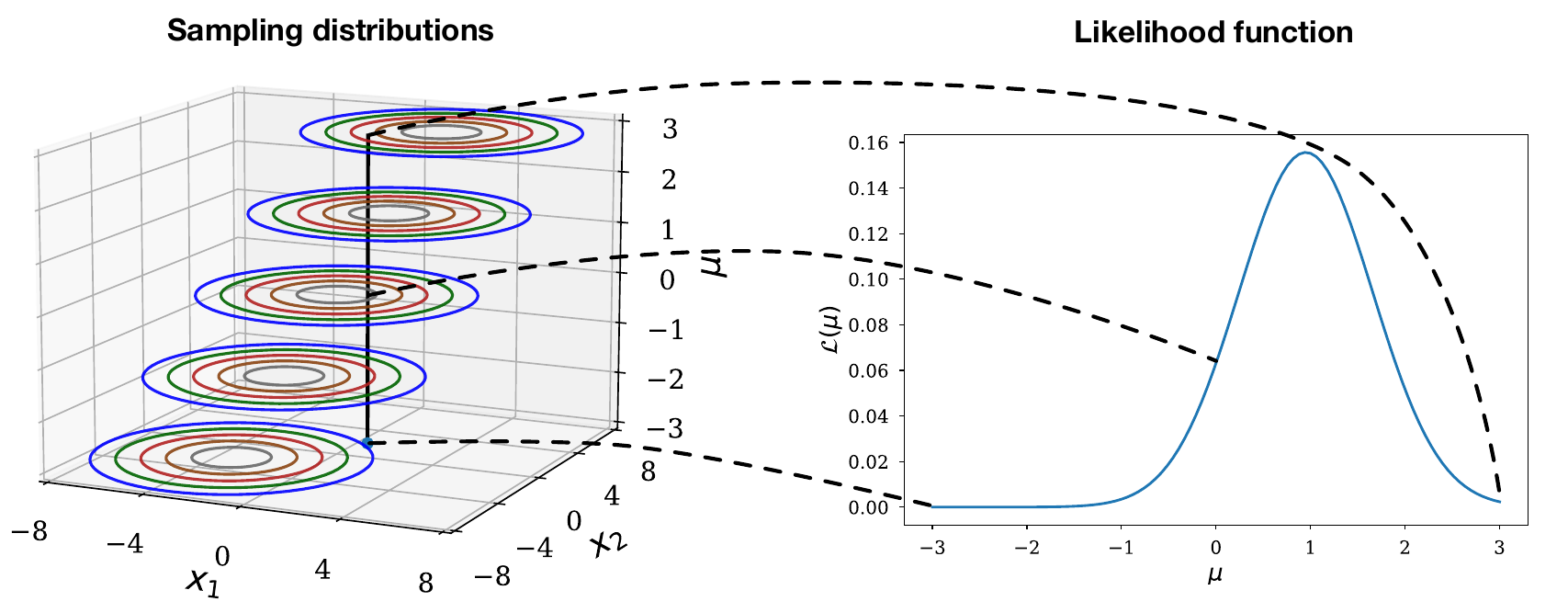}
\end{center}
\caption{The relationship between the collection of sampling distributions comprising a parametric model for 2-dimensional continuous data (left) and the likelihood function (right).}
\label{fig:like}
\end{figure}

\cfig{fig:like} shows a complementary case, where both the sample space and the parameter space are continuous, and the sample space is two-dimensional.
It depicts the relationship between the sampling distributions for a model with a parameter $\mu$ determining predictions for 2-dimensional data $(x_1,x_2)$.
Here $\mu$ is the common mean of a two independent normal distributions for $(x_1,x_2)$ (with known standard deviation $\sigma$, specified in $\ctxt$), so
\be
p(x_1, x_2|\mu,\ctxt) = 
  \frac{1}{\sigma\sqrt{2\pi}} \exp\left[-\frac{(x_1-\mu)^2}{2\sigma^2}\right] 
  \times \frac{1}{\sigma\sqrt{2\pi}} 
  \exp\left[-\frac{(x_2-\mu)^2}{2\sigma^2}\right].
\label{bvn-mu}
\ee
For each choice of $\mu$ (vertical axis), the model specifies a 2-dimensional normalized \pdf\ for $(x_1,x_2)$, depicted via sets of contours in $(x_1,x_2)$ space for each of five values of $\mu$.
Once data are observed, we fix $(x_1,x_2)$ to those observed values and evaluate the sampling distribution as a function only of $\mu$.
This corresponds to slicing through the sampling distributions along the vertical black line.
(See \citet{B80-BayesSciModeling} for a similar figure showing how the posterior PDF relates to the joint distribution for data and parameters.)

The likelihood function quantifies how well each of the candidate sampling distributions---labeled by the parameter $\mu$ in the last example---predicts the observed values of the data, $(x_1,x_2)$.
Note that a statistical model specifies sampling distributions \emph{for data}, and a likelihood function \emph{for parameters}.
``Likelihood for the data'' is incorrect usage---it entirely misses the point of introducing the likelihood terminology.
Sir Ronald Fisher introduced the term specifically to make a distinction between probability and this measure of prediction quality:
\begin{quote}
If we need a word to characterise this
relative property of different values of $p$ [a parameter], I suggest that we may speak without confusion of the likelihood of one value of $p$ being thrice the likelihood of another, bearing always in mind that \emph{likelihood is not here used loosely as a synonym of probability}, but simply to express the relative frequencies with which such values of the hypothetical quantity $p$ would in fact yield the observed sample (\citet{F22-MathFound}, emphasis added).
\end{quote}
Alas, colloquially ``probability'' and ``likelihood'' \emph{are} synonyms, inviting misuse.
The clash between the terminology and colloquial use is unfortunate; ``predictive'' or ``prognostic'' might have been better names for this function.

Fisher goes on:
\begin{quote}
Likelihood also {differs from probability} in that it is a differential element, and is \emph{incapable of being integrated}: it is assigned to a particular point of the range of variation, not to a particular element [interval] (\citet{F22-MathFound}, emphasis added).
\end{quote}
He used stronger language earlier:
``\ldots the integration with respect to $m$ [a parameter] is illegitimate and has no definite meaning\dots'' (\citet{F12-FitFreqLike}, a paper written when he was a third-year undergraduate at Cambridge!).

The posterior \pdf\ is closely related to the likelihood function, obtained simply by multiplying by a prior \pdf\ and normalizing:
\be
p(\theta|\Dobs) \propto p(\theta) \times \like(\theta).
\label{PostPriorLike}
\ee
The posterior \pdf\ is a probability distribution over $\theta$, whose integrals are meaningful.





\section{Priors are not (merely) penalties}

The distinction between likelihood and probability points to the dual role of prior probabilities in Bayesian inference.
Bayesian tutorials typically focus on priors as adjusters of the likelihood to account for base rates, parameter constraints (such as positivity), or results from prior experiments.
This is indeed an important capability.
But more fundamentally, the role of the prior is to ``flip the conditional,'' that is, to get us from likelihood to probability.
With a probability distribution in hand, we are then able to use the LTP to address a myriad of questions involving consideration of composite hypotheses.
In this section we elaborate on the role of priors in Bayesian inference, beyond the usually emphasized role of ``modulating'' the likelihood to account for prior information.

\subsection{Intensive vs.\ extensive quantities in inference}

For an audience of physical scientists, a thermodynamic analogy may be illuminating.
In thermodynamics, temperature is an \emph{intensive} quantity.
It is meaningful to talk about the temperature, $T(x)$, at a point in space, $x$, but not about the ``total temperature'' for a volume of space; temperature does not add or integrate across space.
Heat, on the other hand, is an \emph{extensive} quantity; in mathematical parlance, it may be described by a \emph{measure} (a cumulative mapping from sets or regions, rather than points, to a real number).
The two quantities are related; the heat in a volume $V$ is given by
\be
Q = \int_V dx\, [\rho(x)c(x)]\, T(x),
\label{heat-temp}
\ee
where $\rho(x)$ is the matter density and $c(x)$ is the specific heat capacity.  The product $\rho c$ is extensive, and serves as a kind of conversion factor between temperature and heat.

Priors play a similar role in Bayesian inference, not just modulating the likelihood function, but, more fundamentally, converting \emph{intensive} likelihood to \emph{extensive} probability.
In thermodynamics, a region with a high temperature may have a small amount of heat if its volume is small, or if, despite having a large volume, $\rho c$ is small throughout.  
In Bayesian inference, a region of parameter space with high likelihood may have a small probability if its volume is small, or if the prior \pdf\ assigns low probability to the region.
Approaches that rely exclusively on likelihood focus on particular hypotheses that are ``hot'' (e.g., with maximum likelihood), while Bayesian inference focuses on sets of hypotheses with the most ``heat'' (i.e., with large probability).

\subsection{Typical vs.\ optimal parameter values}

Frequentist statistics includes \emph{penalized maximum likelihood} methods that multiply the likelihood by a penalty function, $r(\theta)$ (e.g., a regularizer), and then locate the maximum of the product:
\be
\tilde{\theta} \equiv \arg\max_\theta r(\theta)\,\like(\theta).
\ee
The penalty function shifts the location of the maximum.
It \emph{looks} like a prior, in that it multiplies the likelihood function, but because Bayesian calculations integrate rather than maximize over parameter space, the prior can do much more than shift the location of the mode (the location of the peak of the posterior) with respect to the maximum likelihood location.

A number of concepts from different areas of mathematics can help us understand the consequences of having a (summable/integrable) measure and not just a preference ordering over a parameter space:
\bi
\item \emph{Curses of dimensionality} observed in high-dimensional geometry, perhaps best known for their impact on numerical computation.
An well-known example is the accumulation of volume close to the edges of a hypercube as dimension grows.
\item \emph{Concentration of measure} in measure theory, showing that volume in a high-dimensional space (with a kind of symmetry) can concentrate in a surprising way in a small region of the space.
\item \emph{Typical sets} in information theory, indicating that the mode of a discrete \pmf\ or a continuous \pdf\ can be highly unrepresentative of typical samples from the distribution \citep{CT06-ElemInfoTheory,B17-IntroHMC}.
\ei
We will give a sense of how these notions can help us understand the impact of integrating vs.\ optimizing with two simple examples.

Consider $N=1000$ flips of a coin with $\alpha \equiv P(\text{heads}) = 0.8$, with the goal being to predict the number of heads.
The sequence of outcomes with the highest probability is the sequence with \emph{1000 heads}.
Despite that, students of probability theory (and probably many less educated!) would predict $\approx 800 \pm 28$ heads.
How can we reconcile this discrepancy?
The sequence with 1000 heads is 
$\left(\frac{\alpha}{1-\alpha}\right)^{200} \approx 2\times 10^{120}$ times more probable than any sequence with 800 heads; that is, it's \emph{hugely} more probable.
But there are \emph{many, many} sequences with $\approx 800$ heads ($\sim 10^{217}$ within one standard deviation of 800 heads); the probability of that so-called \emph{typical set} is $\sim 1$.
A typical random sequence of flips does not at all resemble the most probable sequence.

This phenomenon is not unique to discrete spaces.
Consider now curve fitting with $N=1000$ $(x_i,y_i)$ datapoints, modeled with $y_i = f(x_i) + \eps_i$, for some known parametric function $f(x)$, and with standard normal errors $\eps_i$ (denoted collectively as $\vec{\eps}$).
The most probable sample has $\eps_i = 0$ \emph{for all 1000 samples}.
But those familiar with weighted least squares (minimum $\chi^2$) fitting know that we expect $\chi^2$, the sum of squared residuals, to be $\approx N \pm \sqrt{2N}$ (i.e., $\chi^2$ should be nearly equal to the ``degrees of freedom'').
Note that $\chi^2$ is just the squared distance of the (estimated) $\vec{\eps}$ vector from the origin.
The familiar degrees of freedom result is telling us that a set of random $\vec{\eps}$ vectors will lie in a thin shell far from the origin when $N$ is large, even though the probability density for $\vec{\eps}$ is much higher near the origin than in that shell (one of the ``curses of high dimensionality'').
This is true even though the \pdf\ for the errors has its peak at the origin.
The \emph{density} of points is maximized at the origin, but \emph{volume} grows so quickly with radius in high dimensions that a typical random vector will point far from the origin.

\begin{figure}
\begin{center}
\includegraphics[width=0.515\textwidth]{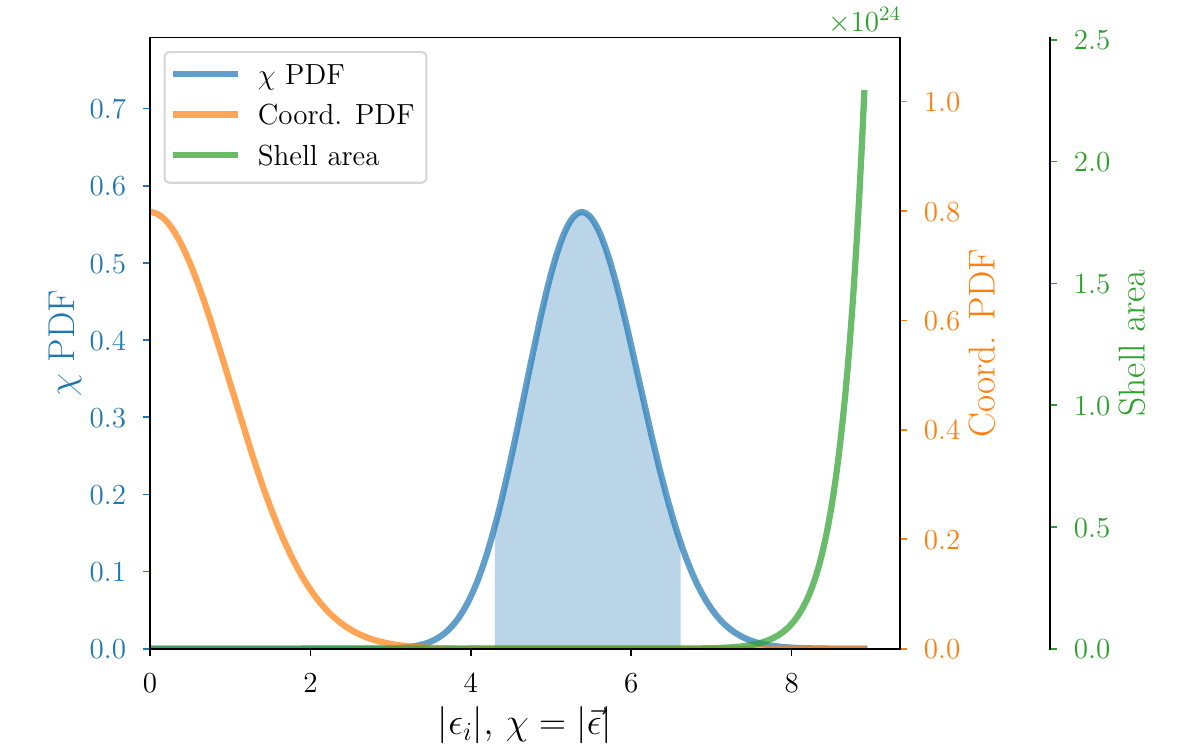}
\includegraphics[width=0.45\textwidth]{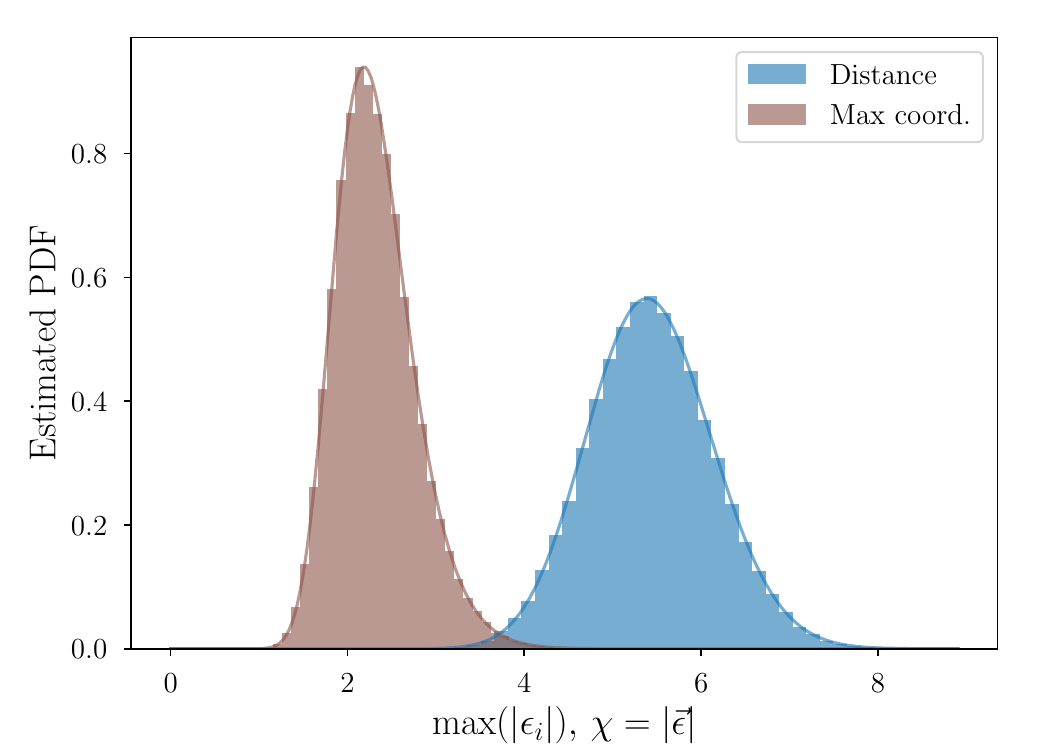}
\end{center}
\caption{Elucidation of the nature of the typical set for a 30-dimensional standard normal distribution with coordinates $\epsilon_i$ and coordinate vector distance $\chi = |\vec{\epsilon}|$ (the square root of $\chi^2$).
\emph{Left:}
Orange curve (against first right ordinate) shows the PDF for a single coordinate value, with highest density at the origin.
Green curve (against the second right ordinate) shows the area of a 29-dimensional hyperspherical surface as a function of $\chi$.
Blue curve (against the left ordinate) shows the PDF for $\chi$, proportional to the product of the area curve and the 30 normal coordinate PDFs.
There is 90\% probability for $\chi$ lying in the shaded region; i.e., typical $\vec{\epsilon}$ draws from a 30-dimensional standard normal will lie in a shell $\approx 5.4$ standard deviations from the origin.
\emph{Right:}
PDFs for $\chi$ (blue) and $\max |\epsilon_i|$ (brown), the maximum size of one of the coordinates, estimated with histograms of $10^5$ draws of $\vec{\epsilon}$ vectors (curves show analytical PDFs).
The blue histogram empirically duplicates the $\chi$ PDF from the left panel.
}
\label{fig:ChiDistn}
\end{figure}

The left panel of \cfig{fig:ChiDistn} shows this quantitatively for $N=30$.
The orange curve shows the PDF of the absolute value of a single $\eps_i$ coordinate, peaking at the origin.
The green curve shows how the surface area of a 29-dimensional sphere (and thus volume in a shell at constant distance) grows with distance from the origin, $\chi$ (the square root of $\chi^2$).
The blue curve shows the PDF for distance, proportional to the product of the area curve and the 30 normal coordinate PDFs.
The shaded region shows the central 90\% region of the $\chi$ PDF; the lengths of typical $\vec{\eps}$ vectors will lie in this region.
That is, typical points in the 30-dimensional $\eps_i$ space lie in a shell about 5.4 standard deviations from the origin.
The right panel of \cfig{fig:ChiDistn} highlights that such intuitively appealing statements do not fully capture the truly non-intuitive nature of distributions in high-dimensional spaces.
We drew $10^5$ samples from a 30-dimensional standard normal, computed the lengths of those vectors, and formed a histogram estimate of the lengths as the blue histogram.
It empirically duplicates the $\chi$ PDF from the left panel.
The brown histogram shows the distribution of the maximum absolute value of the coordinates in each vector.
It reveals that no \emph{single} coordinate value in this large sample ever gets a value near the typical values of $\chi$; put differently, none of the $10^5$ 30-dimensional vectors is close to a coordinate axis.
In the low-dimensional world where our intuition has been trained, these distributions strongly overlap.


The lesson of these examples is that for high-dimensional spaces, the mode of a distribution is likely to be very atypical, in the sense that random draws from the distribution are unlikely to resemble the mode (in terms of the values of the PMF or PDF for the draws).
These examples looked a properties of familiar sampling distributions, but the math does not care if we call a variable ``data'' or ``parameter''---the notion of typical sets applies to distributions over large parameter spaces as well as to distributions over large sample spaces.

\subsection{Example: Histogram-based density estimation}

As an example of these ideas at work in a parameter estimation setting, consider estimating a PDF for some continuous observable quantity via \emph{histogram} data---say, galaxy fluxes grouped into a few dozen flux bins (a ``number-counts'' or ``$\log N$ -- $\log S$'' number-size distribution).
We may model the histogram counts with a \emph{multinomial distribution}, with probabilities $f_k$ for objects to fall in each of $K$ bins ($\sum_k f_k = 1$).
The likelihood function associated with a set of observed counts $n_k$ is then
\be
\like(f_1, \ldots, f_K) = 
  \frac{N!}{\prod_k n_k!} \prod_k f_k^{n_k},
\label{multinom}
\ee
where $N = \sum_k n_k$ is the total number of objects counted.
It is straightforward to show that the maximum likelihood estimates (MLEs) are $\hat f_k = n_k/N$, an intuitively appealing result.

For a Bayesian analysis, a tempting choice of prior, intending to be uninformative about the $f_i$ parameters, is a flat or uniform prior over the simplex $\sum_k f_k = 1$; intuitively, that would seem to be noncommittal.
What does such a distribution say about our prior expectations for $\{f_i\}$?

\begin{figure}
\begin{center}
\includegraphics[width=0.45\textwidth]{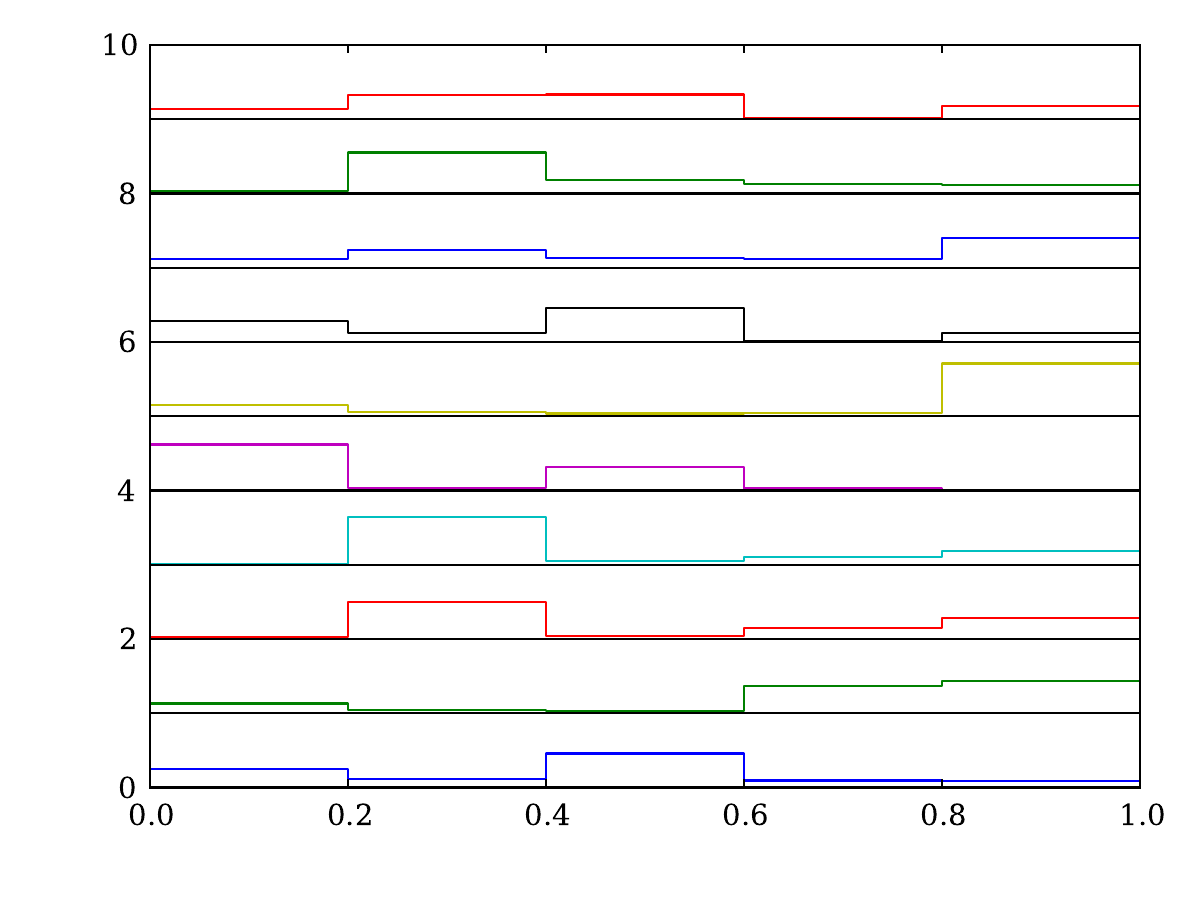} \quad
\includegraphics[width=0.45\textwidth]{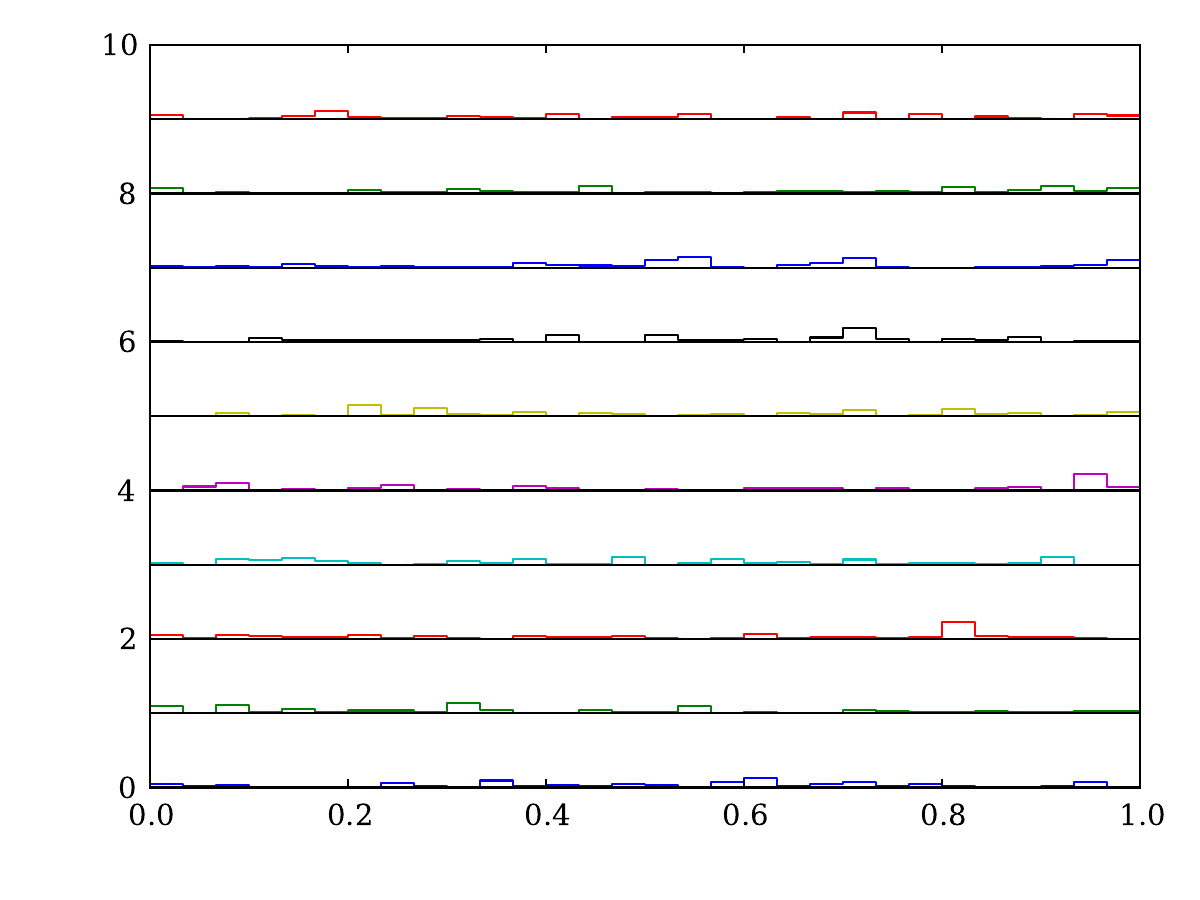}
\end{center}
\caption{10 samples of $f_i$ vectors from a flat prior distribution on the probability simplex, plotted as distributions for a quantity binned over the unit interval.
\emph{Left:} Samples from a distribution for $K=5$ bins.
\emph{Right:} Samples from a distribution for $K=30$ bins.}
\label{fig:DirFlat}
\end{figure}

\cfig{fig:DirFlat} tries to build insight into this question by looking at random samples from the prior.
The left panel shows a stack of 10 random $f_i$ vectors from a flat prior for $K=5$ bins.
The samples show a reasonable amount of diversity, with some samples appearing roughly flat, and some samples having one or more large-$f_i$ bins.
The right panel shows another stack of 10 random vectors, but now for $K=30$.
Perhaps surprisingly, there is now little diversity; all of the samples are nearly flat.


A bit of analytical work confirms the visual impression (see \citealt{FKG10-Dirichlet} for a tutorial covering the math).
We can compute the marginal \pdf\ for a particular $f_k$ (marginalizing over all of the other $f_k$ values).
For $K=2$, this marginal \pdf\ is flat over the unit interval, but as $K$ increases, it increasingly concentrates near $f_k = 0$, with a skew distribution whose expectation value is $1/K$ and whose variance becomes increasingly small.
The random draws and analytical work are telling us that, as dimension grows, most of the volume of a simplex is in the region where all of the $f_k$ are nearly equal to each other, near the middle of the simplex.
(Notably, this is different from the better-known ``curse of dimensionality'' behavior of hypercubes, where volume accumulates at the boundary.)

How might we fix this?
Consider the family of \emph{symmetric Dirichlet priors}, of the form
\be
p(f_1,\ldots,f_K) \propto \delta\left(1-\sum_k f_k\right) \prod_k f_k^{\alpha-1}
\label{SymDir}
\ee
(the $\delta$-function imposes the normalization constraint on the $f_k$ parameters).
Note that the $f_k$ dependence is a product of powers of each $f_k$---the same form of dependence that is in the likelihood function, \ceqn{multinom}.
This prior has a single (scalar) ``hyperparameter,'' $\alpha > 0$, that we can experiment with to control its behavior (it is called the concentration parameter or the flattening parameter for the symmetric Dirichlet).
For $\alpha=1$, the symmetric Dirichlet becomes a flat (uniform) prior.
For other values, the net effect of the prior on the likelihood function is to add $\alpha-1$ to each count, $n_k$; crudely speaking, the prior is adding $\alpha-1$ ``a priori counts'' to the likelihood function.
Finally, this prior is analytically tractable, and produces analytically tractable posteriors.
(This is thanks to \emph{conjugacy}---a Dirichlet prior (symmetric or not), multiplying the multinomial likelihood function, produces a Dirichlet posterior.)

How to set $\alpha$?
One appealing idea recognizes that the binning can be done in different ways, and seeks  some kind of consistency across different binning choices, e.g., different choices of $K$.
Consider in particular \emph{aggregation consistency}:
devise a rule for choosing $\alpha$ such that applying that rule for $K=10$, say, and then aggregating adjacent pairs of bins to produce a 5-bin model, yields the same prior as one would get by applying the $\alpha$ rule directly for $K=5$.
It is not hard to show that setting $\alpha = C/K$ for any constant $C$ satisfies aggregation consistency (for any $K$ and any kind of aggregated binning), while a constant value---like $\alpha = 1$, corresponding to the uniform prior used for \cfig{fig:DirFlat}---does not.
(\citealt{P47-Priors} appears to have offered the first argument along these lines.)


\begin{figure}
\begin{center}
\includegraphics[width=0.45\textwidth]{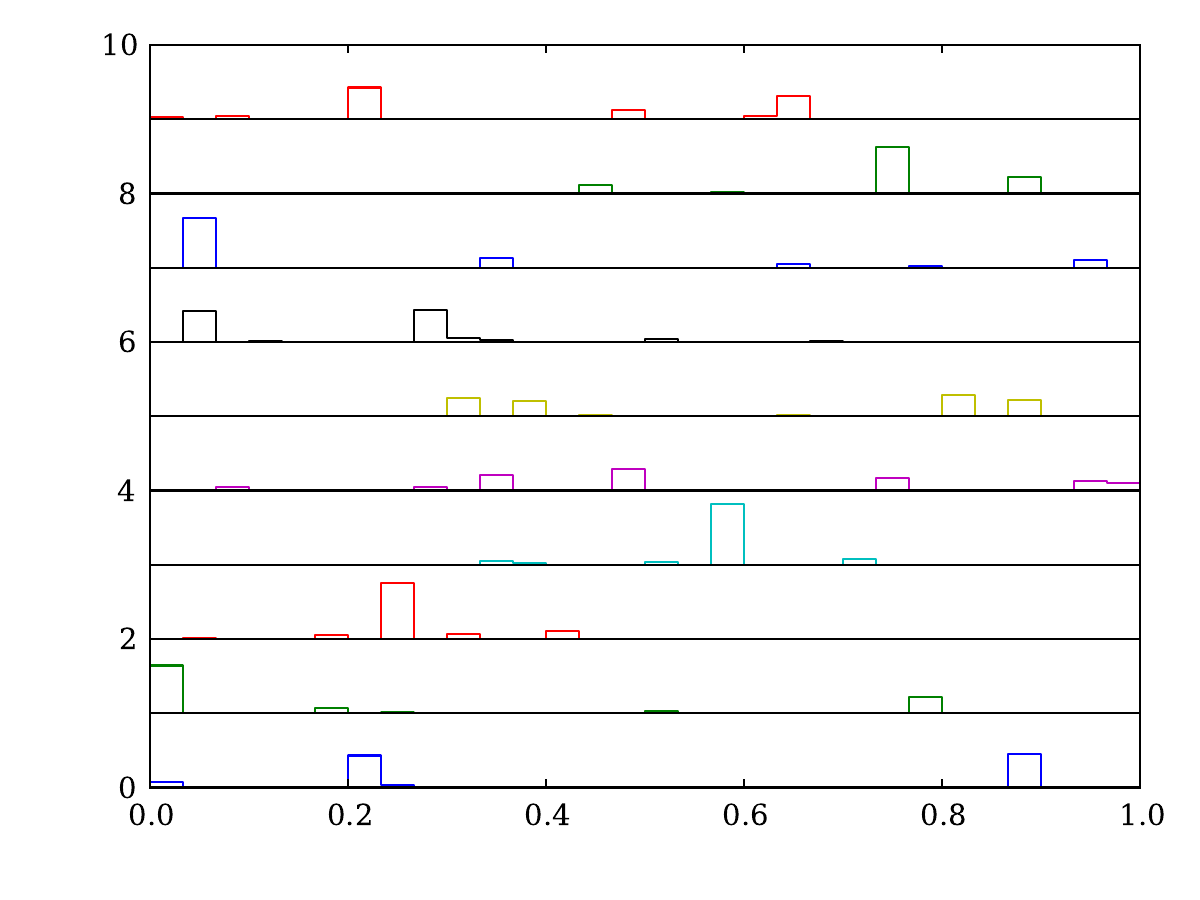}
\end{center}
\caption{As in \cfig{fig:DirFlat}, but with samples drawn from an aggregation-consistent symmetric Dirichlet distribution for $K=30$ bins, with $\alpha = 2/K$.}
\label{fig:DirAgg}
\end{figure}

\cfig{fig:DirAgg} shows samples from an aggregation-consistent prior with $C=2$, so $\alpha = 2/K$.
For $K=2$ bins, this becomes a flat prior, which we saw has acceptable behavior for small $K$.
The figure shows prior samples for $K=30$, to be compared with the right panel in \cfig{fig:DirFlat}.
The prior samples no longer are all flat; in most samples, one or more bins may have significantly larger values of $f_i$ than other bins.

We have focused on the prior; what is the impact on the posterior?
For estimating \emph{all} $K$ bin heights $f_i$ jointly, the impact is \emph{negligible} if there are at least a few counts per bin; the new prior effectively subtracts $C/K -1$ from each $n_k$ counts value, which is not much of a change compared to use of the uniform prior.
However, if we are interested in just one or few bins, and so will marginalize over many other $f_k$ parameters, the difference can be dramatic.
Finally, if we want to do \emph{model comparison}, comparing models with different numbers of bins via their marginal likelihoods (thus marginalizing over all $f_i$ parameters), then the two priors can lead to very different results.
See \citet{L12-BS9} for further discussion and examples.

A lesson of this section is that we should be cautious and skeptical of our intuition when attempting inference in high-dimensional parameter spaces (\cite{G+04-NonparamCMB} have urged similar caution from a frequentist perspective).
Prior probabilities let us get something integrable from likelihood functions, enabling diverse useful calculations when we are interested in composite hypotheses.
But the ``extensive'' (summable) nature of prior and posterior distributions means that parameter space volume effects can have surprising impacts on inference.
If we think of priors merely as modulating or penalizing the likelihood function, it would seem that a uniform prior would not divert attention from the peak of the likelihood function.
But the notion of typical sets, and the behavior of uniform priors in multinomial inference, warn us that \emph{even flat priors may divert attention from the maximum likelihood region when the parameter space is high-dimensional}.
On the one hand, this is a warning that optimization can be misleading; it can focus attention on parts of parameter space that are very atypical.
On the other hand, it is also a warning against relying on low-dimension intuition, like equating prior uniformity with being noncommittal or uninformative.
A practical lesson is to \emph{always look at samples from the prior}, particularly when working in high-dimensional parameter spaces.

\section{Nuisance parameters: Marginalization vs.\ optimization}
\label{sec:marg}

Of the many tasks using the LTP listed in \SS~\ref{sec:LTP}, one of the most important is marginalizing over nuisance parameters.
It is very common (in our experience, essentially ubiquitous) for a data modeling problem to require introducing parameters that are not of direct scientific interest, but that are necessary for describing how the data relate to the quantities of interest.
Estimation of the interesting quantities has to somehow account for the uncertainty in the uninteresting ``nuisance'' quantities.
In this section we elaborate on how marginalization over nuisance parameters accomplishes this, in particular the role of parameter space volume in the behavior of marginal posterior PDFs.

A frequently-arising nuisance parameter setting is analysis of measurements contaminated by \emph{uncertain backgrounds or foregrounds}.
A somewhat more subtle but also prevalent setting is estimating population distributions (density estimation) or correlations and scaling laws for a population (regression), when the properties of members of the population are measured with uncertainty---so-called \emph{measurement error problems} (statisticians use this term for density estimation with noisy measurements, or regression with noise in \emph{both} the measured predictors/covariates ($X$s) and the dependent/response variables ($Y$s); see \citealt{C+06-MsmtErr}).
Examples of density estimation with measurement error (and possibly also with selection effects) include:
estimating  luminosity functions of star, galaxy, and transient object populations;
estimating joint distributions of exoplanet properties such as orbital period and planet size;
estimating distributions of minor planet sizes based on brightness measurements; 
and estimating the distribution of binary black hole system parameters from gravitational wave measurements of masses and spins.
Examples of regression with measurement error include
estimating correlations between galaxy properties (as in the Tully-Fisher and fundamental plane ``luminosity indicator'' relations, or gas density--star formation rate relationships like the Kennicutt-Schmidt relation),
and calibrating the Type~Ia supernova luminosity vs. light curve shape relation.

In a Bayesian treatment of such measurement error problems (and in many frequentist formulations), the unknown true properties for each object formally appear as latent parameters, estimated with uncertainty.
For inference about a population as a whole, the latent parameters are nuisance parameters, and their uncertainty must be propagated into population inferences.
Importantly, in measurement error problems the number of latent parameters grows with sample size; as a result, even small inaccuracies in accounting for measurement error per-object can accumulate and lead to incorrect population-level inferences \citet{L04-SourceUncert}.
\emph{Hierarchical Bayesian (HB) models} handle measurement error by marginalizing over the latent parameters.
Some entry points into the rapidly growing literature on HB demographic modeling in astronomy include: \citet{L04-SourceUncert,K12-MsmtErrModelsAstro,M12-HBSNIa,L13-BackwardLook,AH13-MsmtErrScalingRltns,H+18-ExoRateHBABC,LH19-MultilevelHBCosPop-TLaMPP,MFG19-TLaMPP,V20-HBSlxnEffectsTut}.

\subsection{Understanding marginalization}

\begin{figure}
\begin{center}
\includegraphics[width=\textwidth]{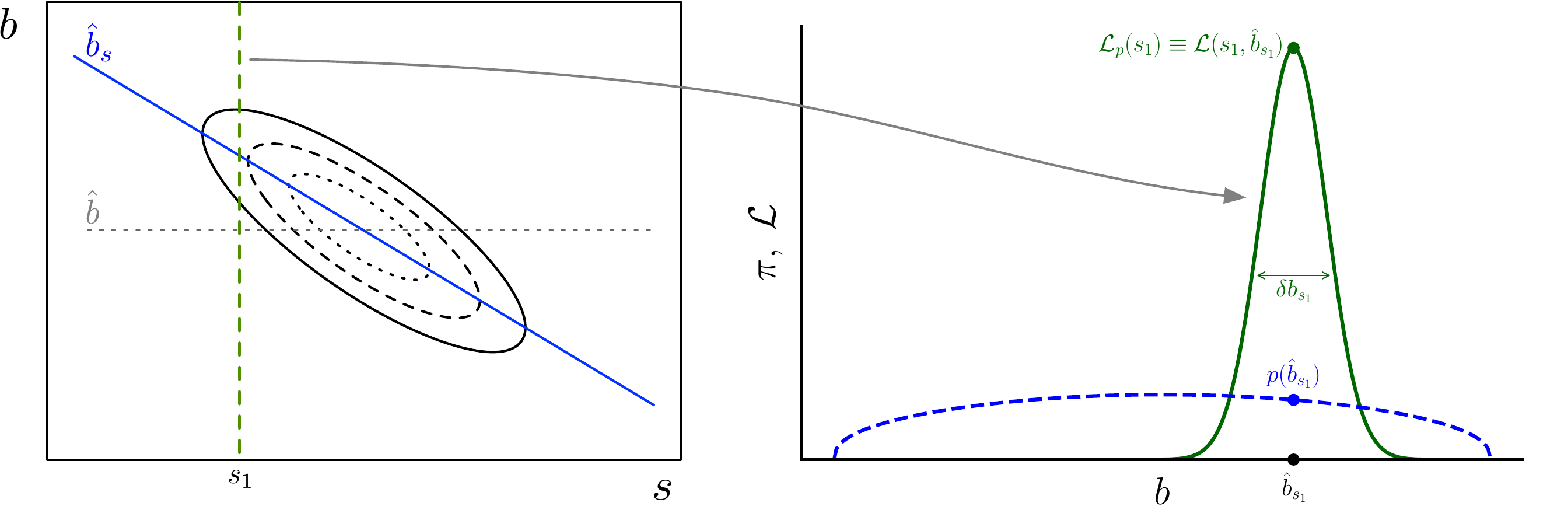}
\end{center}
\caption{\emph{Left:} An example joint likelihood function for $(s,b)$ in a signal-plus-background setting with on/off data that produces a bivariate Gaussian likelihood function.
Gray dotted horizontal line shows $\hat{b}$, the global MLE for $b$.
Blue line shows $\hat{b}_s$, the MLE for $b$ when $s$ is given; it crosses contours of the joint likelihood function where they become vertical.
Vertical green dashed line indicates slice of the joint likelihood function at $s=s_1$ displayed in right panel.
\emph{Right:} Ingredients for approximating the marginal likelihood function for $s$.
Green curve shows a slice of the joint likelihood function at $s_1$, as a function of $b$.
Blue dashed curve shows the (conditional) prior PDF for $b$.
See text for symbols.
}
\label{fig:nuis-binorm}
\end{figure}

Given the prevalence and importance of nuisance parameter problems in astronomy, it is worth building insight into how marginalization accounts for nuisance parameter uncertainty.
To do so, we will consider the signal-plus-background setting, with interesting signal parameter, $s$, and background nuisance parameter, $b$.
We suppose the available data provides useful information about $b$, so the joint posterior PDF, $p(s,b|\Dobs,\ctxt)$, exhibits some dependence between $s$ and $b$.
For example, $\Dobs$ may include independent data about $b$ from an off-source measurement at a location away from a candidate source in an image analysis setting.
In such ``on/off'' settings, the on-source data provide a noisy measurement of $s+b$, and the off-source data provide a separate noisy measurement of $b$.
The joint likelihood function for $(s,b)$ then typically resembles the left panel of \cfig{fig:nuis-binorm}, exhibiting a negative correlation (the larger $b$ may be, the smaller $s$ is likely to be).

Specifying the value of one parameter in a multiparameter problem is a composite hypothesis: specifying just $s$ in a problem with parameters $(s,b)$ corresponds to saying that, for the given $s$, \emph{some} parameter choice in the set $\{(s,b): b \in [b_l, b_u]\}$ holds, though we don't know which (here $[b_l, b_u]$ denotes the range of allowed $b$ values, e.g., $[0,\infty]$).

To summarize implications for $s$, accounting for $b$ uncertainty,
we compute the marginal posterior PDF for $s$,
\be
\begin{split}
p(s|\Dobs,\ctxt) &= \int db\; p(s,b|\Dobs,\ctxt)\\
  &\propto \int db\; p(s,b|\ctxt)\, \like(s,b),\\
\end{split}
\label{bmarg}
\ee
where the second line uses the numerator of BT, with likelihood function $\like(s,b)$ and a joint prior PDF $p(s,b|\ctxt)$.

Now factor that joint prior as the prior for $s$, $p(s|\ctxt)$, times a conditional prior for $b$ given $s$, $p(b|s,\ctxt)$.
In many settings there may be no contextual information linking $b$ and $s$ a priori, making this factor independent of $s$; to simplify the notation, we assume such independence, so the $b$ prior factor is simply $p(b|\ctxt)$.

Since the marginalization integral is over $b$, the $s$ prior can be taken out of the integral:
\be
\begin{split}
p(s|\Dobs)
  &\propto p(s) \int db\; p(b)\, \like(s,b)  \scond{\ctxt}\\
  &\propto p(s)\, \like_m(s)
\end{split}
\label{bmarg-fac}
\ee
where we have defined the \emph{marginal likelihood function} for $s$ as
\be
\like_m(s)
   \equiv \int db\; p(b)\, \like(s,b)
\label{mlike-s}
\ee
(where we suppress $\ctxt$ henceforth to simplify the notation).
Since the integration underlying marginalization is isolated to the marginal likelihood function, we will focus henceforth on $\like_m(s)$.

To build insight into what marginalization accomplishes, we approximate the integral over $b$ in two steps (the ingredients for this approximation are depicted in the right panel of \cfig{fig:nuis-binorm}):
\begin{enumerate}
\item Assume the data are informative, so the prior PDF for $b$ is nearly constant in the region where the integrand in \ceqn{mlike-s} has most of its area.  This lets us pull the $p(b)$ factor out of the integral, setting it equal to its value at the location of the peak of the sliced likelihood function.
\item Approximate the remaining integral of $\like(s,b)$ over $b$---the area under the likelihood function in the nuisance parameter dimension---as the product of the height of the likelihood function and some convenient measure of its width (in the $b$ dimension). 
\end{enumerate}
We need some notation to implement these steps.
Let $\hat{b}_s$ denote the value of $b$ that maximizes the likelihood function when $s$ is fixed, i.e., along a vertical slice on the left panel of \cfig{fig:nuis-binorm} (the depicted slice has $s=s_1$):
\be
\hat{b}_s = \arg \max_b \like(s,b).
\label{hat-b-def}
\ee
Graphically, $\hat{b}_s$ as a function of $s$ traces the points where tangents to the contours of $\like(s,b)$ become vertical.
The uninformative prior assumption of Step~1 implies
\be
\int db\; p(b|s)\, \like(s,b) 
  \;\approx\; p(\hat{b}_s|s) \int db\; \like(s,b).
\label{Lmarg-approx-1}
\ee
The integral over $b$ gives the area under the likelihood function curve.
Let $\delta b_s$ denote some convenient measure of the width of the sliced likelihood function.
Approximate the integral as the product of the height and width of the sliced likelihood function,
\be
\int db\; \like(s,b) \approx \like(s,\hat{b}_s)\;\delta b_s
\label{like-area}
\ee
The first factor is the joint likelihood function maximized over $b$ for each $s$.
This is a function solely of $s$; it is called the \emph{profile likelihood function} for $s$,
\be
\like_p(s) \equiv \like(s,\hat{b}_s).
\label{like-p-def}
\ee
It corresponds to looking at the 3D $\like(s,b)$ surface along the $b$ direction, producing a 2D projection or profile of the surface.

Assembling these results, we have this helpful approximation for the marginal likelihood function for $s$:
\be
\like_m(s) \approx p(\hat{b}_s)\,\like_p(s)\, \delta b_s.
\label{Lmarg-approx}
\ee
The last two factors are the data-dependent factors.
Much of the work done by marginalization is captured in the profile likelihood factor.
A naive approach to handling a nuisance parameter would be to fix it at its best-fit value.
This would correspond to slicing $\like(s,b)$ horizontally along the gray dotted line (through the global MLE point, $(\hat{s},\hat{b})$).
The profile likelihood function instead slices $\like(s,b)$, accounting for the fact that a choice of $s$ implies a different ``best'' choice for $b$ than the global MLE.
But the marginal likelihood function includes the additional factor, $\delta b_s$, that accounts for how the \emph{uncertainty} in the nuisance parameter may change as a function of the interesting parameter.
Essentially, marginalization is saying to start by profiling (rather than just using the best-fit nuisance parameter), but then adjust the profile likelihood, weighting regions where the nuisance parameter uncertainty is large more heavily than regions where it is small.

\cfig{fig:nuis-binorm} depicts a joint likelihood function that is a bivariate Gaussian function.
Recall that a bivariate Gaussian may be written as the product of a Gaussian for one variable (say, $s$ here) and a Gaussian for the second variable with mean related linearly to the first one (the $\hat b_s$ line in the figure), and \emph{constant} standard deviation in the second dimension.
This implies that $\delta b_s$ is constant with respect to $s$ here: the marginal likelihood function is proportional to the profile likelihood function.
But that is a special property of multivariate Gaussians that does not hold in general.

\begin{figure}
\begin{center}
\includegraphics[width=\textwidth]{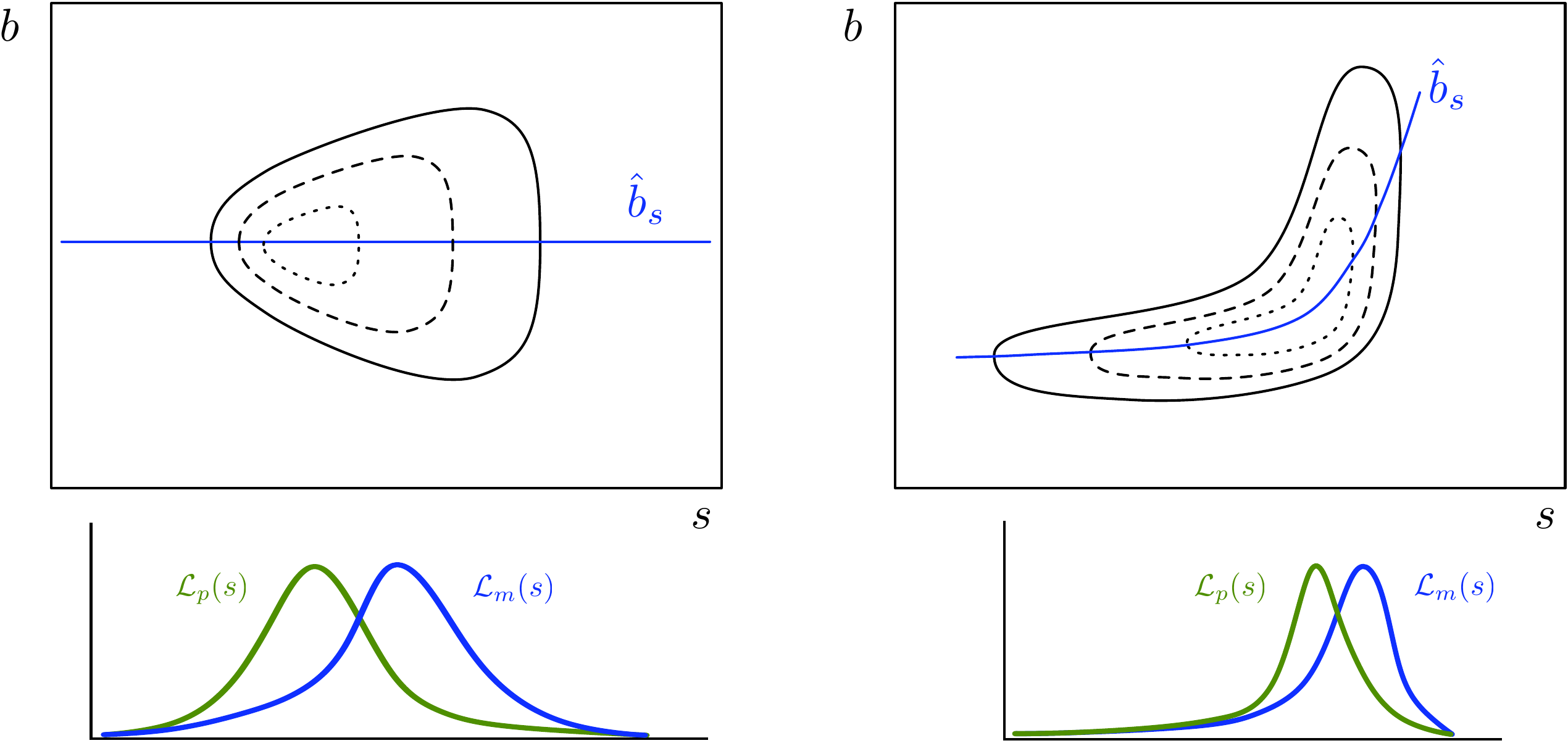}
\end{center}
\caption{Illustration of two-parameter likelihood function geometries where marginalization and profile likelihood may lead to significantly different inferences.
Top panels show joint likelihood functions for parameter of interest $s$ and nuisance parameter $b$.
Blue curves show $\hat{b}_s$, the MLE for $b$ when $s$ is given, the path defining the profile likelihood function.
Bottom panels show profile (green) and marginal (blue) likelihood functions for $s$.
\emph{Left:} A likelihood function with ``flaring'' contours, with the nuisance parameter uncertainty varying significantly as a function of the interesting parameter.
\emph{Right:} A likelihood function with ``banana-shaped'' contours, with nuisance parameter uncertainty varying because of the curvature of the profile path.
}
\label{fig:flare-banana}
\end{figure}

\cfig{fig:flare-banana} shows two scenarios where marginal and profile likelihood functions may noticeably differ (here $s$ and $b$ denote generic interesting and nuisance parameters).
The top left panel shows a ``flaring'' two-variable likelihood function, with the uncertainty in $b$ changing significantly as a function of $s$.
The bottom left panel indicates how the flaring shifts the marginal likelihood function away from the joint peak, reflecting that the larger $b$ uncertainty to the right makes it more likely the true parameter values are to the right of the MLE point.
The top right panel shows a ``banana-shaped'' two-variable likelihood function.
These arise commonly in settings where predictions depend on the parameters via power laws (so predictions vary more weakly along a curve than orthogonal to it).
Here the $\delta b_s$ weighting of the profile likelihood function arises because as the likelihood contours tilt upward with increasing $s$, the uncertainty in $b$ increases.

\subsection{Measurement error problems---handling many nuisance parameters}

It is often the case that the nuisance parameter volume correction is noticeable in a single measurement, but not hugely significant; it may shift the profile likelihood function by only a fraction of its width.
But the effect tends to be in a similar direction for similar measurements.
When combining information across many measurements, the systematic effect from the nuisance parameter volume factor can accumulate, and ignoring it can corrupt aggregated inferences, such as demographic inferences.

\begin{figure}
\begin{center}
\includegraphics[width=\textwidth]{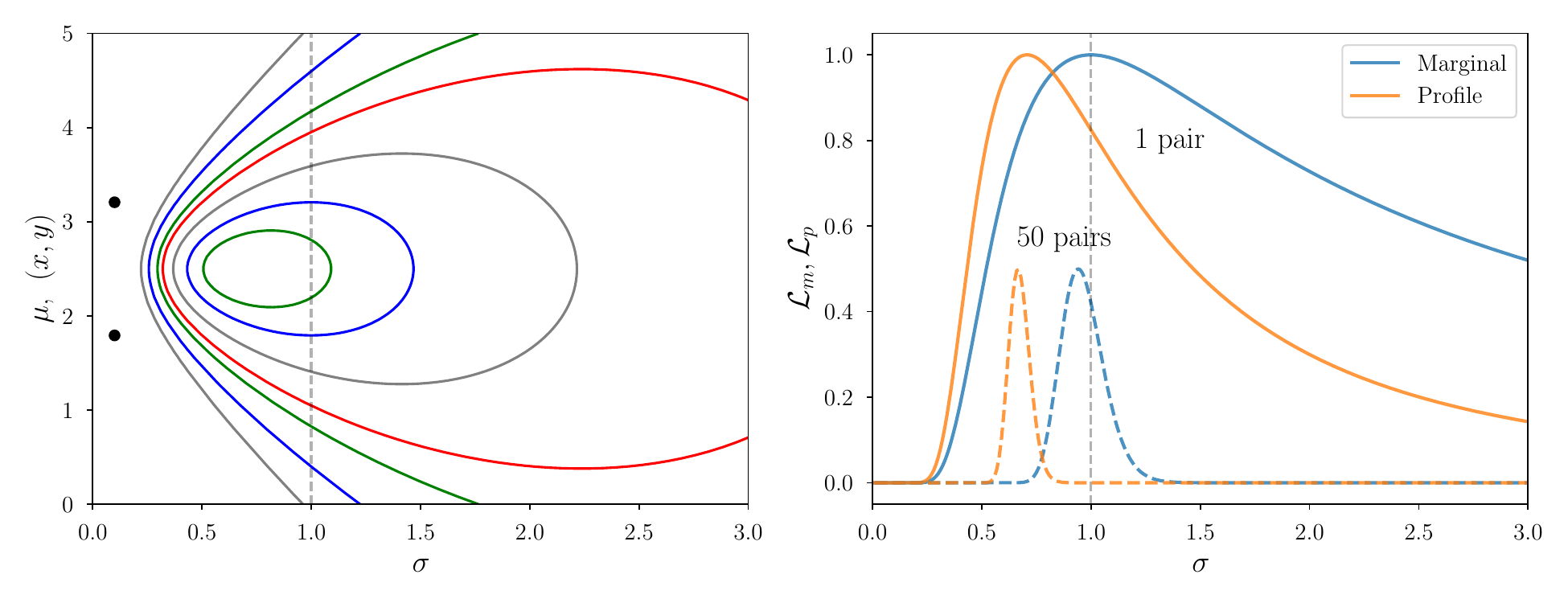}
\end{center}
\caption{Joint, marginal, and profile likelihood functions for the normal pairs (Neyman-Scott) problem.
\emph{Left:} Joint likelihood function for the standard deviation $\sigma$, and the mean $\mu$ of a normal distribution, inferred using a pair of samples (depicted as dots adjacent to the $\mu$ axis).
Vertical dashed line shows the true value of $\sigma$ for the ensemble of pairs of which this pair is a member.
Contours bound approximate 25\%, 50\%, 75\%, 90\%, 95\%, 99\%, and 99.9\% confidence regions.
\emph{Right:} Likelihood functions for $\sigma$ based on a single pair (solid curves, scaled to unit height) or a set of measurements of 50 pairs (dashed curves, scaled to height 0.5 for visibility).
Blue curves are marginal likelihood functions, orange curves are profile likelihood functions.
Vertical dashed line shows the true value of $\sigma$ for the ensemble of pairs.
}
\label{fig:NeymanScott}
\end{figure}

\cfig{fig:NeymanScott} illustrates this with a somewhat contrived but illuminating example.%
\footnote{Statisticians know this problem as the \emph{Neyman-Scott problem}, a well-known counterexample to maximum likelihood estimation \citep{NS48-NSProblem}.
Neyman and Scott did important work on statistics in astronomy.
Although the Neyman-Scott problem was first described in an econometrics journal, they posed it in terms of measuring ``some physical constant such as the radial velocity of a star or the velocity of light.''
They noted the connection to measurement error problems, and described several settings in astronomy where the issues they addressed potentially arise.
Our discussion here expands on a shorter treatment in \cite{L04-SourceUncert}.
}
For each of a set of $N$ objects, we make a pair of measurements of some object property, $\mu_i$ for object $i$, with an instrument that provides measurements with additive Gaussian noise with zero mean and fixed but \emph{unknown} standard deviation, $\sigma$.
Denote the pair of measurements by $x_i$ and $y_i$.
For object $i$, there is a likelihood function for its $\mu_i$ and the shared $\sigma$ parameter that is the product of two normal distributions,
\be
\ell_i(\mu_i,\sigma) =
  \frac{1}{\sigma\sqrt{2\pi}} \exp\left[-\frac{(x_i-\mu_i)^2}{2\sigma^2}\right] 
  \times \frac{1}{\sigma\sqrt{2\pi}} 
  \exp\left[-\frac{(y_i-\mu_i)^2}{2\sigma^2}\right].
\label{NS-memlike}
\ee
From a single pair of measurements, there is substantial uncertainty in both $\mu_i$ and $\sigma$.
The left panel in \cfig{fig:NeymanScott} displays this; it shows contours of $\ell_i(\mu_i,\sigma)$ for the pair of measurements depicted as dots along the ordinate.
Notably, this is an example of a flaring likelihood function, which we see is not an unusual phenomenon; it is relevant even for the familiar normal distribution when there is uncertainty, not just in its mean, but also in its standard deviation.

We should be able to improve inference by pooling information across many pairs of measurements.
Each pair brings in more information about $\sigma$, letting us calibrate the noise level of the instrument.
That combination of $\sigma$ information should then let us measure each $\mu_i$ more precisely, as if $\sigma$ became known.
The right panel of \cfig{fig:NeymanScott} shows ingredients for inferring $\sigma$ from the pooled data.
The solid curves show the profile (orange) and marginal (blue) likelihood functions from a single pair of measurements (for the marginal, a uniform prior was used for $\mu_i$).
The behavior illustrated in \cfig{fig:flare-banana} is apparent here; the peak of the marginal is shifted to the right of the peak of the profile curve.
Moreover, the marginal curve is considerably broader.
That said, the two curves have substantial overlap.
From a single pair measurement, inferences based on the two curves would differ noticeably, but perhaps not strongly.

The dashed curves in the right panel of \cfig{fig:NeymanScott} show what happens when we accumulate information across multiple paired measurements (here drawn from a broad, uniform distribution, though the findings described here do not depend on this choice).
Formally, we do this by computing a likelihood function for the full dataset that is the product of the likelihood functions for the pairs,
\be
\like(\vec{\mu}, \sigma) = \prod_{i=1}^N \ell_i(\mu_i,\sigma),
\label{NS-like}
\ee
where $\vec{\mu}$ denotes the collection of $\mu_i$ parameters.
We can again summarize the implications for $\sigma$ by profiling or marginalizing over $\mu_i$---but now over all $N$ $\mu_i$ parameters.
The dashed curves show the resulting profile and marginal likelihood functions.
Both are becoming usefully narrow.
But the profile likelihood function is converging \emph{away} from the true value ($\sigma=1$ here).
One can show that the MLE is in fact an \emph{inconsistent} estimator for $\sigma$ in this setting---it converges to the wrong value as $N\rightarrow\infty$ (this remains the case even if more than two measurements are made per object, provided the number of per-object measurements is finite).
In the first (frequentist) treatment of this problem, Neyman and Scott \citep{NS48-NSProblem} noted that a useful estimate for $\sigma$ could be obtained by averaging unbiased moment-based point estimates of $\sigma$ from the pairs.
But they commented that ``This is undoubtedly true but beside the point'' that such problems expose a problem with maximum likelihood estimation that has no general solution.
(Bayesian methods were uncommon in statistics at that time, and they did not explore marginalization.)

\begin{figure}
\begin{center}
\includegraphics[trim={.4cm .5cm .4cm 0},clip,width=0.75\textwidth]{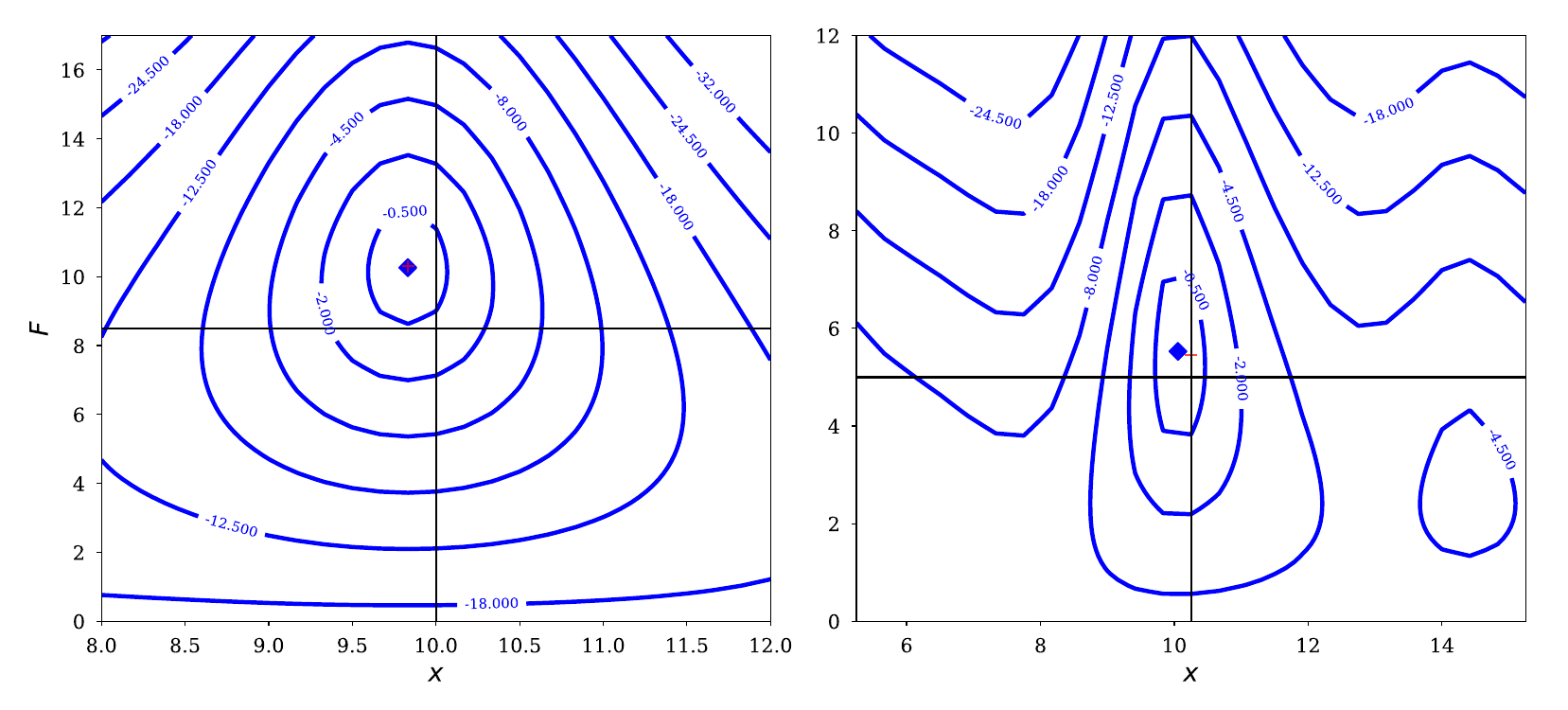}
\end{center}
\caption{Likelihood functions as a function of flux, $F$, and one position coordinate, $x$ for dim point sources in an astronomical image.
Crosshairs show the true $(x,F)$; diamonds shows maximum likelihood estimates.
Contours have nominal $1\sigma$ spacing.
\emph{Left:} Likelihood for a source that would just pass a $6\sigma$ detection threshhold.
\emph{Left:} Likelihood for a source that would just pass a $3\sigma$ detection threshhold.
}
\label{fig:2DImage}
\end{figure}

To make the point that such problems are not uncommon in astronomy, 
\cfig{fig:2DImage} shows likelihood functions from simulated data for isolated \emph{dim} point sources observed with a CCD camera with a critically-sampled Gaussian PSF (the calculations are based on the description of the Hyper Suprime-Cam pipeline in \citealt{B+18-HSC}).
The left panel is for a $\approx 6\sigma$ source whose true position is in the middle of a pixel.
Contours are in the space of flux, $F$, and one of the direction coordinates, $x$ (position on the CCD; the other coordinate is fixed at its best-fit value).
The contours exhibit flaring; they are are nearly symmetric in $x$, but are very asymmetric in $F$, showing the \emph{significant flux-direction dependence}: positions away from the best-fit position, in either direction, imply lower flux estimates.
The growth of area in the contours with decreasing flux means that \emph{the marginal likelihood for flux gets shifted downward from the joint best-fit flux}.
The shift is at the percent level, much smaller than in the paired measurement problem above.
But, as in that problem, the shift is \emph{systematic}, in the same direction and of similar scale for all $6\sigma$ sources; it impacts accuracy (bias), more than than precision, and would dominate statistical errors in analyses averaging over a few hundred sources or more.
The right panel shows likelihood contours for a $\approx 3\sigma$, sub-threshold measurement of a source with a quarter-pixel offset.
It shows that the likelihood function rapidly becomes more complicated for sources dimmer than $\approx 5\sigma$, amplifying the need to thoroughly account for nuisance parameter uncertainty when studying the flux distribution of dim sources (e.g., via forced photometry).

For those requiring a frequentist solution to measurement error problems, there is a large and growing literature presenting numerous techniques specialized to specific problem settings.
From a Bayesian perspective, specialized inference techniques are unnecessary; marginalization over latent parameters handles such problems generally and flexibly (though the resulting computational challenges can demand nontrivial computational algorithms specialized to different settings).
Notably, even statisticians who generally favor frequentist methods recommend Bayesian approaches that marginalize over latent parameters in complicated settings, particularly for nonlinear modeling of data with heteroscedastic measurement error (i.e., with measurements that have differing standard deviations); see \cite{C+06-MsmtErr}.
Astronomical data commonly has heteroscedastic measurement error, and the ability of marginalization to flexibly accommodate this complication has been a strong motivation for the spread of Bayesian methods for measurement error problems in astronomy.



\section{Systematic error and model misspecification}

A well-known aphorism attributed to George Box motivates our final marginalization application area: ``All models are wrong, but some are useful'' (\citealt{BD87-ModelBuilding}; see also \citealt{B76-SciStat}).
In some data analysis settings, we may be largely secure about our model, but still want to explore the possibility that inference may be corrupted by potential influences omitted from the model.
In other settings we may knowingly adopt an approximate model, and want to account for the lack of fidelity to the data when we make inferences about salient features of the model.
Here we briefly describe the role composite hypotheses and marginalization can play to improve uncertainty quantification in such settings.
We consider two settings:  first, where systematic error arises due to model selection uncertainty (i.e., whether to include terms for anticipated corrupting effects), and second, where a model aims to account only for salient features of a phenomenon, and we want to account for uncertainty in deliberately unmodeled details.

\subsection{Systematic error and model selection uncertainty}

A fairly common way astronomers try to account for systematic error from identified potential corrupting effects is to use a classical null hypothesis significance test (NHST) to see if there is significant evidence in the data for the effects.
A statistic is devised to test for the presence of a potential effect; if the null hypothesis of no effect is not rejected, analysis proceeds assuming no effect is present.
A significant issue with this approach is that failure to reject a null hypotheses need not correspond to a strong preference for the null over the alternative (see, e.g., \citealt{B03-FisherTesting,WL16-ASAPValues,G+16-PValInterp,WSL19-BeyondP}).
Ideally we would like to take into account potential systematic effects, weighted in a way that measures the strength of evidence for or against their presence.

Marginal likelihoods can provide such a weighting (see the brief description of \emph{model averaging} above).
However, marginal likelihoods typically are more sensitive to the choice of prior than are parameter estimates---in particular, to the ranges of the parameter spaces considered for rival models---and one must be cautious about using them to account for systematic error.

An example of systematic error quantification using marginal likelihoods and Bayes factors is the work of \citet{DLW00-SNIa} (DLW00) exploring the potential impact of systematic error from source evolution on early analyses of the evidence for dark energy from measurements of Type~Ia supernova (SN~Ia) light curves (``SN cosmology'').
The dark energy discovery papers found no significant evidence for SN~Ia source evolution (dependence of light curve properties on redshift), and so assumed there was \emph{exactly zero evolution} in their analyses.
DLW00 computed Bayes factors indicating the data were equivocal regarding the presence of evolution.
An important aspect of the calculation was consideration of a variety of priors for parameters in evolution models, to ensure the findings were robust.
Allowing for plausibly small levels of evolution, and marginalizing over its impact, significantly weakened the strength of the evidence for dark energy, \emph{unless one assumed a flat cosmology}, which was not justified at the time.
Fortunately, within a few years accumulating evidence from other cosmological phenomena honed in on flat cosmologies.
From the perspective of the DLW00 systematic error analysis, it was only in the context of this later evidence that the discovery of dark energy became secure.

\subsection{Systematic error, overdispersion, and model discrepancy}

In astronomy we often knowingly use models that capture only the \emph{salient features} of a phenomenon.
This is particularly true when first exploring a frontier area, where learning such features can provide significant insight, even if the salient feature model overlooks details.
Examples include counting pulses in gamma-ray burst prompt light curves (where we may care only about the number of pulses and their time scales, not every minor wiggle), or modeling luminosity functions with broken power laws (where we would like to estimate a power law slope and a break luminosity, irrespective of minor bumps in the distribution).
Simply adopting the salient feature model and ignoring model misspecification can corrupt inferences, in particular by providing artificially tight constraints on the salient model parameters.

One approach to handling this is to simply inflate sampling distributions.
For example, in the statistics literature analyzing count data, it is common to adopt negative binomial (NB) rather than Poisson distributions at the outset, since the NB distribution has an extra parameter that can be used to overdisperse the predictive distribution for counts, and it includes the Poisson distribution as a special case (see, e.g., \citealt{H11-NegBinom}, and \cite{dS+15-GLM-NB} for an application in astronomy; see the Appendix for a brief discussion of the NB distribution and its use for modeling overdispersion).
\citet{B23-SysErrChiSqr} has adapted this idea to account for systematic error in astronomical data analysis of Poisson data in the regime where $\chi^2$ fitting is nominally accurate, devising a new \emph{overdispersed $\chi^2$ distribution}.

We are developing complementary methods for salient feature modeling of Poisson count and point process data that are closely tied to use of NB distributions.
But we do not adopt overdispersed distributions outright.
In contrast to most applications of the Poisson distribution in the statistics literature, in many astrophysical applications there is a strong physical basis for adopting a Poisson distribution as the description of the repeated-sampling variability expected in the data---the so-called \emph{aleatoric uncertainty} (from the Latin \emph{aleator} for ``dice player'').
We use overdispersion to reflect a combination of aleatoric and epistemic sources of uncertainty (with the epistemic component capturing the systematic errors that may not vary randomly across replications).


\begin{figure}
\begin{center}
\includegraphics[width=0.55\textwidth]{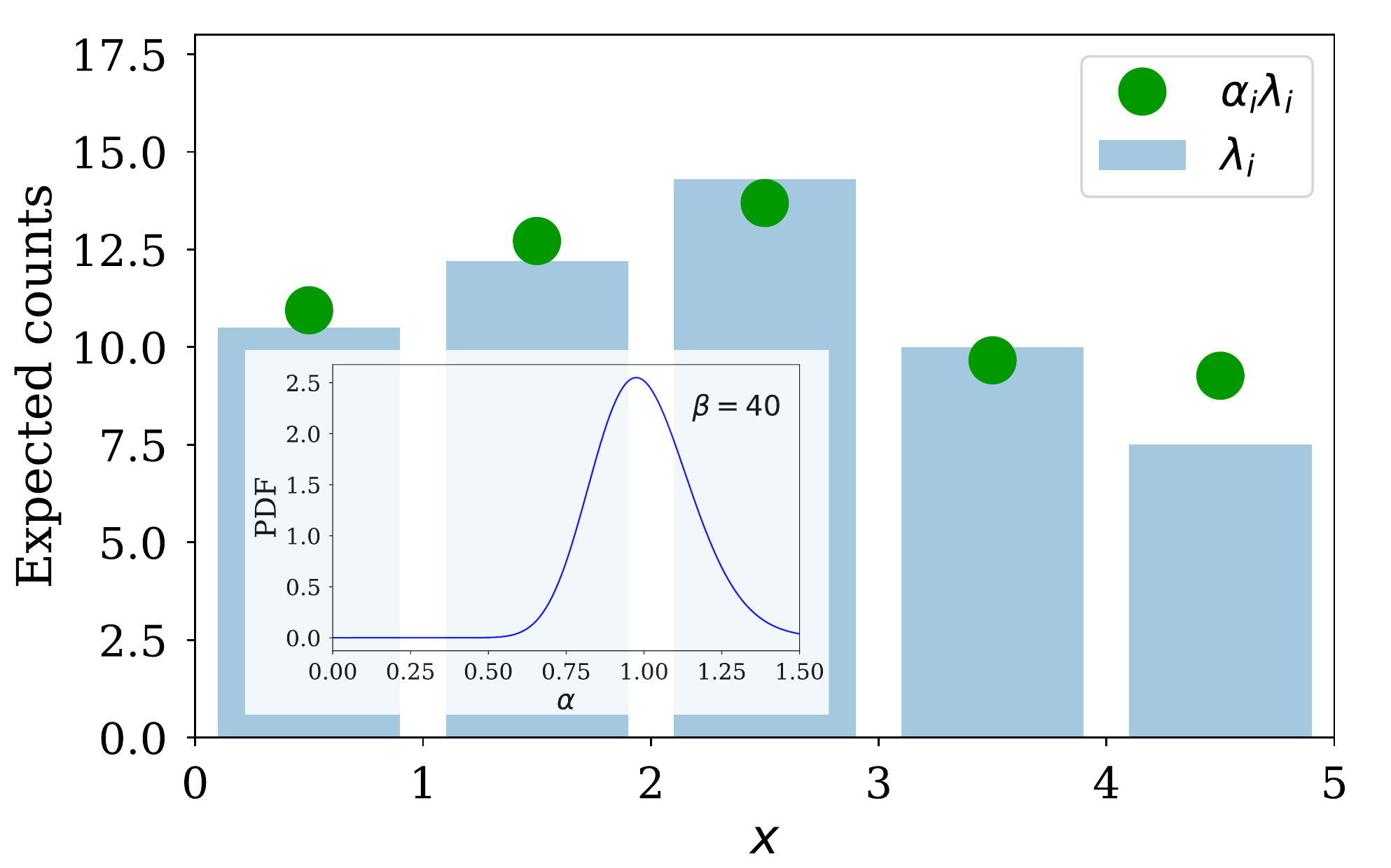}
\end{center}
\caption{Depiction of the multiplicative discrepancy approach.
Blue histogram shows the Poisson distribution PMF from a salient feature model.
Green dots depict the PMF resulting from multiplying the Poisson expectation values by draws from a gamma distribution discrepancy process.
Inset shows the unit-mean gamma distribution for the discrepancy factors.}
\label{fig:Salient}
\end{figure}

We construct overdispersed distributions via representations that tie them to the underlying aleatoric sampling distribution via a plausible \emph{discrepancy mechanism} (adapting the notion of additive discrepancy functions from Gaussian process emulation of computer models; see the Appendix).
We consider the actual intensity function (event rate) for the Poisson process governing the data to be the product of the parametric salient model rate that we wish to estimate, and a nonnegative discrepancy factor.
\cfig{fig:Salient} illustrates the construction for a model for photon counts from a time-varying source, binned in time, used to estimate parameters of a salient light curve model.
The blue histogram depicts the predictions of a Poisson model for the photon counts in bins, as expectation values, $\lambda_i$.
These are from the salient feature model, an idealized description of the true light curve that captures key features of scientific interest (e.g., location, duration, amplitude).
To predict the observed counts in a bin, $n_i$, we multiply each intensity, $\lambda_i$, by a random \emph{discrepancy factor}, $\alpha_i$, drawn from a distribution with positive support and unit mean, shown in the inset.
We use a gamma distribution for $\alpha_i$; with the mean fixed to one, there is only a single parameter, the shape parameter, $\beta$, controlling the variance.
In the fitting process, we marginalize over all of the $\alpha_i$ factors, and treat $\beta$ as an uncertain parameter, estimated from the data.

The choice of a gamma distribution enables \emph{analytic} marginalization over the discrepancy factors, producing a predictive \pmf\ for the counts that is NB, with a variance that may be inflated with respect to that of a Poisson distribution.
(See the Appendix for further details.)
More deeply, the gamma choice lets us introduce discrepancy in a manner that preserves aggregation consistency in a sense analogous to what we described above regarding Dirichlet priors for histograms.
This framework explicitly acknowledges that the actual count variability distribution is expected to be Poisson.
It also points toward generalizations that can apply to point data, and that can account for bin-to-bin correlations in the discrepancy.
Details for specific astrophysical applications will be reported elsewhere.

\section{More than Bayes's theorem}


The theme of this tutorial has been that BT and the LTP are partners in Bayesian inference---arguably unequal partners, with the LTP carrying much of the burden in analysis and computation.
This is because of how common composite hypotheses are in astrophysical data analysis.
We hope we have made a strong case that the role of the LTP and marginalization should be highlighted more prominently in the Bayesian astrostatistics literature, particularly in pedagogical presentations.

Python and R code producing figures from this paper are available at \url{https://github.com/tloredo/MoreThanBT2024-Figures}.

\section*{Appendix: Marginalization in overdispersion and discrepancy models}

\newcommand{\param}{\psi}
\newcommand{\nbd}{\mathcal{N}}

\subsection{Overdispersion and discrepancy for data with additive Gaussian noise}

Consider curve fitting (regression) of noisy measurements, $y_i$, that are functions of known values of a quantity $x_i$---for example, fitting spectral data, with $y_i$ being flux density, and $x_i$ being wavelength ($i=1$ to $N$).
Let $f(x;\param)$ be the model predicting the true value of $y$ as a function of $x$, with parameters $\param$.
The function $f(x;\param)$ may be a straightforward formula (like a black body or power law spectrum), or it may only be computable via numerical computer modeling (e.g., from a radiative transfer code, as used for fitting exoplanet atmosphere spectral data).
Our goal is to estimate the parameters, $\param$.

Suppose the measurements have additive zero-mean Gaussian errors, each with some known error standard deviation, $\sigma_i$.
Then we may write the model as
\be
y_i = f(x_i; \param) + \eps_i; \qquad 
\eps_i \overset{\mathrm{ind}}{\sim} \normal(0,\sigma_i^2),
\label{y-f-norm}
\ee
with $\normal(0,\sigma^2)$ denoting a zero-mean normal distribution with variance $\sigma^2$.
The likelihood function can be computed from the product of $N$ normal PDFs for the $y_i$ measurements,
\be
\like(\param) \propto \frac{1}{\prod_i \sigma_i} \exp\left[-\frac{\chi^2(\param)}{2}\right],
\label{L-chisqr}
\ee
where $\chi^2(\param)$ is the usual weighted sum of squared residuals measure of misfit,
\be
\chi^2(\param) \equiv \sum_i \left( \frac{y_i - f(x_i;\param)}{\sigma_i} \right)^2.
\label{chisqr}
\ee
Note that we have kept the $1/\sigma_i$ factors (from the normal distribution  normalization constants) in \ceqn{L-chisqr}, for reasons that will become apparent.

Parameter inference would multiply the likelihood function by a prior and explore the resulting posterior distribution.
But an important component of Bayesian workflow (and of frequentist modeling) is \emph{model checking}---providing some kind of test of adequacy of the assumed model, typically by measuring how typical or atypical the observed data are from what the model predicts.
Frequentist model checking typically examines the \emph{minimum} value of $\chi^2$.
Bayesian model checking examines the posterior-averaged value of $\chi^2$, to account for parameter uncertainty.
In either case, if that value is too much larger than $N$, that is reason for concern about the model.

What to do then?
One path is to formulate rival $f(x;\param)$ functions and compare their performance to the original one (say, via Bayesian model comparison).
But another way the original model may have failed is that the noise scale may have been misspecified, i.e., the $\sigma_i$ values are wrong.
An informal way astronomers sometimes handle this is to inflate the $\sigma_i$ values until the minimum $\chi^2$ is equal to $N$ (which has the desirable effect of broadening the likelihood function).
From a Bayesian perspective we can implement this by introducing an overdispersion parameter, $\alpha$, and changing the noise distribution model to
\be
\eps_i \overset{\mathrm{ind}}{\sim} \normal(0,\alpha^2\sigma_i^2),
\label{norm-alpha}
\ee
which changes the likelihood function to
\be
\like(\param, \alpha) \propto \frac{1}{\alpha^N \prod_i \sigma_i} \exp\left[-\frac{\chi^2(\param)}{2\alpha^2}\right].
\label{L-chisqr-alpha}
\ee
(We kept the $1/\sigma_i$ factors in \ceqn{L-chisqr} to make sure that the $\alpha$ dependence of the normalization constants of the overdispersed Gaussians shows up in the revised likelihood function.)
We can then introduce a prior for $\alpha$ (e.g., enforcing $\alpha > 0$, and perhaps reflecting that we expect $\alpha$ to be near unity and likely to be above unity) and marginalize over it to estimate $\param$, accounting for uncertainty in the true $\sigma_i$ noise scales.
The marginalization ends up broadening the likelihood function somewhat more than the $\min\chi^2 = N$ approach, because it accounts for uncertainty in $\alpha$ (i.e., in the true noise scale).%
\footnote{The marginalization can be done analytically if one adopts an $\alpha$ prior that is uniform in $\alpha$ or $\log\alpha$, or, more informatively, with priors of the form of an \emph{inverse gamma} PDF, a two-parameter family of distributions that can express a preference for $\alpha$ values near unity, for example.}

This overdispersion approach reflects the hypothesis that the actual noise level is larger than the originally assumed level.
We could potentially test it with repeated observation of the same system, where we expect residuals to vary with the larger noise scale.

An alternative way to interpret too-large $\chi^2$ is that the model, $f(x;\param)$, captures the essential or salient behavior of the system, but that there are small departures from $f$ not captured by the salient model, large enough that we should consider how they may impact estimation of $\param$.
We can express this by writing
\be
y_i = f(x_i; \param) + \delta_i + \eps_i,
\label{y-f-disc}
\ee
where we have added a new \emph{discrepancy} term to the model, $\delta_i$, meant to describe the difference between the actual function predicting $y_i$ and our $f$ model (now explicitly regarded to be approximate).
We need to assign a joint probability distribution to all of the $\delta_i$ values.
A simple choice is to describe our discrepancy uncertainty with independent zero-mean normal distributions for each $\delta_i$, with standard deviations proportional to the noise scales, $s_i = \beta\sigma_i$.
The sum of two independent zero-mean random variables with normal distributions is itself normal, with variance given by the sum of the separate variances.
So the net effect of this discrepancy model is to rescale $\sigma_i$ by $1+\beta$.
If we reparameterize, setting $\alpha = 1+\beta$, then we end up with the same likelihood function as in \ceqn{L-chisqr-alpha}.
Inference can proceed as before.
But the interpretation is different.
Here, $\delta_i$ represents epistemic uncertainty about how the true function predicting $y_i$ relates to the salient but not quite accurate model function $f$.
In this setting, were we to repeatedly observe the same system, we would not expect the residuals to vary with scales larger than $\sigma_i$.
Rather, we would expect the $\delta_i$ to be the same in repeated observations; we would be able to estimate them, revealing how the salient and true models differ.

This kind of discrepancy modeling arises fairly commonly in settings where $f$ is computationally expensive, and must be interpolated from a set of runs from a computer model.
Gaussian process models are flexible interpolators frequently used for this task, including in astronomy (e.g., \citealt{H+12-GPEmuCosmo,C+24-OstrichPCA+GP}).
In such settings, more complex and better-motivated $\delta_i$ models are used than the one we described above.
Commonly, the $\delta_i$ are modeled in a correlated manner, enforcing some level of smoothness in the discrepancy as a function of $x$, using an additional Gaussian process.
This enables analytical marginalization over possible structured discrepancy functions.
\cite{H+04-GPDiscrepancy} provide a tutorial.

\subsection{Overdispersion and discrepancy for Poisson-distributed data}

Many astronomical measurements comprise counts of discrete objects or events, often well-modeled using the Poisson distribution.
This is particularly the case for image, spectral, and time series data that count photons, where the fundamental physics of emission and detection processes often strongly justifies using Poisson distributions.
An interesting question is how the overdispersion and discrepancy ideas can be implemented in Poisson settings.

For concreteness, consider a times series setting, with photon counts binned in time, with $n_i$ counts in bin $i$, and with bins of equal duration, $\Delta$.
Suppose we have a model that predicts the photon counting rate (the Poisson point process \emph{intensity function}), $\lambda(t;\param)$, as a function of time, $t$, and of the model parameters, $\param$.
The likelihood function for $\param$ is given by the product of Poisson distributions for the counts in each bin;
\be
\like(\param) = \prod_i \frac{\left[\lambda(t_i;\param)\Delta\right]^{n_i}}{n_i!}
  \exp\left[-\lambda(t_i;\param)\Delta\right]
\label{L-pois}
\ee
(for simplicity we have assumed the rate does not vary strongly over $\Delta$, so the expected counts is given by the product of the rate at the mid-time of a bin, $t_i$, and the bin duration).
We can estimate $\param$ using this likelihood in a similar manner to use of \ceqn{L-chisqr} for data with additive Gaussian noise.

Now suppose there is evidence of misfit for this model, i.e., that the differences between the observed counts, $n_i$, and the expected counts, $\lambda(t_i;\param)\Delta$ are larger than the Poisson standard deviation, given by the square root of the expected counts.
In some settings, the Poisson assumption may just be a starting point, without a compelling physical justification, and one may seek an alternative PMF that offers more control over the dispersion.
For the Poisson distribution, both the mean and the standard deviation are set by $\lambda$.
The \emph{negative binomial} (NB) distribution is a count distribution that offers separate control of the mean and standard deviation allowing for overdispersion of counting data.
In astronomy, \cite{dS+15-GLM-NB} have used the NB to model counts of globular clusters around a galaxy as a function of galaxy luminosity, bulge velocity dispertion, and other properties.

In problems with binary outcomes (say, success and failure), with success probability $\theta$, the familiar binomial distribution of \ceqn{binom} gives the probability for seeing $n$ successes in $N$ total trials; $N$ is fixed a priori, and $n$ is random.
The NB distribution gives the probability for a related setup where one counts the number of failures, $n$ (redefined here), before a specified target number of successes, $r$, is reached.
The number of failures is random and it can range from $0$ (when all $r$ successes happen at the start of the sequence) to $\infty$.
Thus the total number of trials is random and can range from $n$ to $\infty$).
The NB PMF is
\be
\begin{split}
p(n|\theta,r) 
  &= \frac{(n+r-1)!}{n! (r-1)!}\, (1-\theta)^n \, \theta^r \\
  &\equiv \nbd(n; r, \theta).
\label{p-NB}
\end{split}
\ee
(This is just the binomial probability for seeing $r-1$ successes in $n+(r-1)$ trials, and then seeing a success on the next trial.)
Mathematically, this PMF for $n$ is well-defined even for non-integer (positive) values of $r$ (in which case the factorials must be computed using gamma functions).
The $\theta$ and $r$ parameters enable separate control of the mean ($\mu = r(1-\theta)/\theta$) and standard deviation for $n$.
The NB PMF becomes a Poisson PMF with mean number of counts $\mu$ in the limit where $\theta\rightarrow 1$ with $\mu$ held fixed (corresponding to holding $r = \mu\theta/(1-\theta)$).
Inference of $\param$ in settings with overdispersed counts proceeds using NB, typically reparameterized in terms of $\mu$ (e.g., given by $\lambda(t;\param)\Delta$) and an overdispersion parameter.

How could we introduce the discrepancy idea into modeling Poisson count data?
It turns out the NB distribution is useful for this, arising as follows.
We suppose $\lambda(t;\param)$ is a salient model, not quite specifying the actual Poisson mean.
Write the actual Poisson mean in bin $i$ as the product of the salient mean and a \emph{multiplicative disrepancy factor}, $\alpha_i$ (playing a role analogous to that of $\delta_i$ in the Gaussian setting), so that the likelihood function now takes the form
\be
\like(\param;\{\alpha_i\}) = \prod_i \frac{\left[\alpha_i\lambda(t_i;\param)\Delta\right]^{n_i}}{n_i!}
  \exp\left[-\alpha_i\lambda(t_i;\param)\Delta\right].
\label{L-pois-disc}
\ee
We describe uncertainty in the discrepancy factors via a PDF for the $\alpha_i$ values that has an expectation value of unity, so that the salient model is expected to be accurate in an average sense (over the bins).
A convenient choice is a \emph{gamma distribution} (a power law times an exponential in $\alpha$), with its mean fixed at unity, and a single parameter controlling its width (the spread of the discrepancy factors).
With this choice, we can marginalize over the $\alpha_i$ parameters analytically; this yields a NB distribution for the counts in each bin.
The marginal likelihood function for the rate function parameters $\param$ can be written as
\be
\like_m(\param) = \prod_i \nbd\left(n_i; \; \beta\lambda(t_i;\param)\Delta, \, \frac{\beta}{1+\beta} \right),
\label{L-pois-disc-m}
\ee
where $\beta$ is the parameter controlling the width of the gamma distribution for the discrepancy factors.
With this parameterization, the Poisson standard deviation is inflated by $\sqrt{(1 + 1/\beta)}$.
Here the NB distribution arises, not as a description of overdispersion we would expect to see as excess count variability in repeated observation of the same system, but rather to account for epistemic uncertainty about the discrepancy between the salient feature model, $\lambda(t_i;\param)$, and the true event rate.
This framework is especially appropriate in settings where the physics strongly motivates the Poisson model as a description of aleatoric uncertainty (expected variability), since it presumes there is a latent model that describes the data with the Poisson distribution.

The gamma distribution choice enabled us to analytically marginalize over the discrepancy factors.
But the gamma choice has a more fundamental motivation.
One can imagine rebinning the data (either splitting or aggregating bins).
We would like the discrepancy model to satisfy aggregation consistency (as we did for the Dirichlet example, above):  applying the approach to the original data, and then rebinning, should produce the same inferences as applying the approach to the rebinned data.
The gamma choice enables enforcing this consistency requirement by scaling $\beta$ with the bin size.

We devised this approach for modeling gamma-ray burst time series data for thesis work on new computational algorithms for Bayesian inference done with Duke University PhD student Mary Beth Broadbent \citep{B14-LARKThesis}.
We are currently developing it further, including generalizing the approach to handle ``time-tagged event'' point process data (where the arrival times of individual photons are recorded).

\section*{Conflict of Interest Statement}

The authors declare that the research was conducted in the absence of any commercial or financial relationships that could be construed as a potential conflict of interest.

%

\section*{Funding}
This material is based upon work supported by the National Science Foundation under Grants No.\ DMS-2015386 and AST-2206339 (for TL), and DMS-2015382 (for RW).

\section*{Acknowledgments}
We gratefully acknowledge the organizers and participants of the $\emph{iid2022}$ workshop for helpful interactions on the topic of this paper.
We are also grateful to the issue editor and two reviewers for helpful feedback that significantly improved this paper.

\bibliographystyle{Frontiers-Harvard} 
\bibliography{LW23-IID22Refs}



\end{document}